\documentclass[12pt,english]{article}
\usepackage[T1]{fontenc}
\usepackage[latin1]{inputenc}
\usepackage{geometry}
\usepackage{epsfig}

\usepackage{amsmath}
\usepackage{graphicx}
\usepackage{setspace}
\usepackage{amssymb}
\usepackage{amsfonts}
\makeatletter

\renewcommand{\[}{\begin{equation}}
\renewcommand{\]}{\end{equation}}
\@addtoreset{equation}{section}
  
\usepackage{babel}
\makeatother
\begin{document}
\newcommand{\bra}[1]{\left\langle #1\right|}
\newcommand{\ket}[1]{\left|#1\right\rangle }

\def\bit{\begin{itemize}}
\def\eit{\end{itemize}}
\def\bdes{\begin{description}}
\def\edes{\end{description}}
\def\ben{\begin{equation}}
\def\een{\end{equation}}
\def\bea{\begin{eqnarray}}
\def\eea{\end{eqnarray}}
\def\nn{\nonumber}
\def\ben{\begin{equation}}
\def\een{\end{equation}}
\def\bea{\begin{eqnarray}}
\def\eea{\end{eqnarray}}
\def\R{\rangle}
\def\L{\langle}
\def\half{\frac{1}{2}}
\def\lt{\left}
\def\rt{\right}
\def\ho{\hat{\omega}}
\newcommand{\com}[1]{} 
\renewcommand{\thefootnote}{\fnsymbol{footnote}}
\setcounter{footnote}{1}
\thispagestyle{empty}
\addtocounter{page}{-1}
\begin{flushright}
MCTP-06-15\\
\end{flushright}
\vspace*{1.3cm}
\centerline{\large \bf Integrable Deformations of $\hat{c}=1$ Strings in Flux 
Backgrounds}
\vspace*{1.2cm}
\centerline{\bf Joshua L. Davis\footnote{joshuald@umich.edu}, Finn 
Larsen\footnote{larsenf@umich.edu}, Ross O'Connell\footnote{rcoconne@umich.edu} 
and Diana Vaman\footnote{dvaman@umich.edu}}
\vspace*{0.8cm}
\centerline{\it Michigan Center for Theoretical Physics}
\vspace*{0.2cm}
\centerline{\it University of Michigan, Ann Arbor, MI-48109, U.S.A.}
\vspace*{1cm}
\com{\centerline{\tt ${}^{\clubsuit}$joshuald@umich.edu, 
${}^{\diamondsuit}$larsenf@umich.edu, ${}^{\heartsuit}$rcoconne@umich.edu, 
${}^{\spadesuit}$dvaman@umich.edu}}
\centerline{ \today}
\vspace*{1.5cm}
\begin{abstract}
We study $d=2$ 0A string theory perturbed by tachyon momentum modes in 
backgrounds with non-trivial tachyon condensate and Ramond-Ramond (RR) flux. In 
the matrix model description, we uncover a complexified Toda lattice hierarchy 
constrained by a pair of novel holomorphic string equations. We solve these 
constraints in the classical limit for general RR flux and tachyon condensate. 
Due to the non-holomorphic nature of the tachyon perturbations, the 
transcendental equations which we derive for the string susceptibility are 
manifestly non-holomorphic. We explore the phase structure and critical 
behavior of the theory.

\end{abstract}

\baselineskip=18pt
\newpage

\renewcommand{\thefootnote}{\arabic{footnote}}
\setcounter{footnote}{0}

\section{Introduction}

Two-dimensional string theory has been fruitfully used to explore problems for 
which we lack the technical prowess and/or conceptual framework to address in 
more realistic, higher-dimensional string theories. This has been made possible 
mostly by the matrix models dual to the $d=2$ bosonic (for a review see 
\cite{Klebanov:1991qa,Ginsparg:1993is,Polchinski:1994mb}) and Type 0 strings 
\cite{Douglas:2003up,Takayanagi:2003sm}, which provide non-perturbative 
definitions for those theories. Of interest in the present article are 
Ramond-Ramond flux backgrounds which were recently investigated using the 
matrix dual descriptions in \cite{Maldacena:2005he}. In this work we further 
study non-trivial backgrounds in Type 0A string theory by deforming the theory 
with an integrable set of momentum modes in the presence of both RR flux ($q$) 
and tachyon condensate ($\mu$).

Although the integrable structure of the $d=2$ bosonic string has been 
thoroughly explored (for example, in 
\cite{Dijkgraaf:1992hk,Takasaki:1993at,Takasaki:1994jy,Takasaki:1995fu,Kazakov:2000pm,Kostov:2001wv,Kostov:2002tk,Alexandrov:2002fh,Alexandrov:2003ut,Cremonini:2005sc}), 
such structures for 0A string theory are somewhat less developed. The 
integrable properties of the 0A matrix model were first studied using the Toda 
lattice hierarchy \cite{Ueno:1984} in the early '90s \cite{Nakatsu:1995eg}, 
when it was known as the deformed matrix model \cite{Jevicki:1993zg}. More 
recently this has been discussed by \cite{Yin:2003iv,Park:2004yc}. 
Alternatively, following \cite{Moore:1992ga}, perturbative techniques have been 
utilized to address momentum mode deformations in \cite{PandoZayas:2005tu}. A 
few of the lowest order correlators were computed and a pattern for $\mu=0$ was 
discerned; summing the infinite perturbative series lead to an expression for 
the 0A partition function. As discussed in \cite{Kazakov:2000pm} for the $c=1$ 
matrix model, this is equivalent to solving the string equation constraining 
the integrable Toda lattice hierarchy. However in all of these previous works, 
the authors studied limits where either $q \to 0$ or $\mu \to 0$ to obtain 
tractable results regarding the partition function or string susceptibility.

In contrast, we obtain transcendental equations for the genus zero string 
susceptibility parameterized by non-trivial tachyon condensate, Ramond-Ramond 
flux and momentum mode perturbations simultaneously. To do so we take advantage 
of the complex coordinate $\mu + iq$ identified in \cite{Maldacena:2005he} and 
introduce a complexified Toda lattice hierarchy. This hierarchy is based on 
shifts of both the energy and angular momentum quantum numbers of the 
single-fermion Hilbert space of the dual 0A matrix model description. The 
utilization of the fermion angular momentum as a dynamical variable is 
reminiscent of the proposed non-critical M-theory 
\cite{Horava:2005tt,Horava:2005wm} perspective on 0A string theory, although we 
introduce it for purely mathematical ends. The complex nature of the integrable 
structure provides an additional real constraint (string equation) on the 
operator algebra of the theory, relative to previous integrability analyses. 
This additional constraint allows us to solve for the susceptibility equations 
in the dispersionless, \textit{i.e.} classical, limit.

This paper is organized as follows. Section 2 contains a review of the 
treatment of integrable momentum mode perturbations to the $c=1$ matrix model. 
We introduce our notation and philosophy here. In section 3, we address 
integrable deformations of the 0A matrix model. We show how the system is 
intractable when there is only one string equation as a constraint. We then 
introduce the complexified Toda lattice hierarchy which provides two such 
constraints. We solve this system in the dispersionless limit, obtaining 
equations for the perturbed susceptibility. In section 4 we analyze the 
susceptibility equations, exploring the critical behavior and phase diagram of 
0A string theory. We close with section 5 wherein we include some preliminary 
results on perturbations which take advantage of the holomorphic structure of 
the complexified Toda hierarchy. Finally, a number of technical appendices are 
included to elucidate points made within the text as well as providing an 
alternate derivation of the complex string equations.

While discussing the bosonic string and the $c=1$ matrix model, our units are 
such that $\alpha^\prime =1$. When discussing the Type 0 strings, we will use 
$\alpha^\prime=\half$ units.

\section{Integrable Perturbations of the $c=1$ Matrix Model}\label{c1int}

Before turning to integrable deformations of 0A string theory, we will now 
provide a short review of such deformations in the $c=1$ model, first studied 
in \cite{Dijkgraaf:1992hk}. The purpose of this detour is to introduce our 
logic and notation in an example where we closely follow previous analyses 
before moving on to a treatment of 0A where we will depart significantly from 
prior work by other authors.

\subsection{Chiral Quantization and the Energy Representation }

The $c=1$ matrix model is usefully described by a system of non-interacting 
fermions. The operator algebra of these fermions is given by
\begin{eqnarray}
\left[\hat{x}_{+},\hat{x}_{-}\right] & = & i,\nn\\
\left[\hat{x}_{\pm},\hat{\epsilon}\right] & = & \pm i\hat{x}_{\pm},\nn\\
\left\{ \hat{x}_{+},\hat{x}_{-}\right\}  & = & 2 
\hat{\epsilon}.\label{c=1algebra}
\end{eqnarray}
where $\hat\epsilon$ is the energy and $\hat{x}_\pm = \frac{\hat{p} \pm 
\hat{x}}{\sqrt{2}}$ are light-cone coordinates in the single-particle phase 
space. The commutator of $\hat{x}_\pm$ with $\hat{\epsilon}$ indicates that 
these operators have simple expressions in the energy basis, $\hat{x}_\pm \sim 
e^{\pm i \partial_\epsilon}$. To produce the correct commutator and 
anti-commutator between $\hat{x}_+$ and $\hat{x}_-$ requires the addition of 
certain dressing phases. Suitable energy representations which reproduce all of 
the relations (\ref{c=1algebra}) are
\ben
\lt[\hat{x}_\pm \rt]_\epsilon =\pm 
S^{\mp1/2}\left({\epsilon}\right)\ho^{\pm1}S^{\pm1/2}\left({\epsilon}\right) 
\label{xpm-energy}
\een
where $\ho$ is the shift operator
\ben
\ho  \equiv e^{i \partial_\epsilon}, \label{shift}
\een
and $S(\epsilon)$ is
\ben
S(\epsilon) = e^{-i \pi /4} \sqrt{\frac{\Gamma \lt(\half - i \epsilon 
\rt)}{\Gamma \lt(\half + i\epsilon\rt)}} \equiv e^{i \phi_0 (\epsilon) },
\een
which can be understood as the $\epsilon \to -\infty$ scattering amplitude for 
the fermions.\footnote{See Appendix A for more details on light-cone 
quantization and calculation of $S(\epsilon)$.}

Some comments on the representations above are in order. Although we use the 
term ``operator'' to describe $\ho$, as well as using a hat, it should be made 
clear that in this text we will use it exclusively in the energy basis as a 
shorthand for the derivative \eqref{shift}. Thus, expressions containing $\ho$ 
will not be treated as operator statements but rather as basis specific 
statements. One could, of course, adopt the alternative viewpoint that $\ho$ is 
an operator and \eqref{shift} is its energy representation, but we will not do 
so.

Additionally, we wish to mention that the energy basis for the $\hat{x}_\pm$ 
operators is actually somewhat subtle and depends on the sign of $\epsilon$. In 
studying the $c=1$ matrix model, one considers only fermions with negative 
energy, which in perturbation theory are localized in one of the quadrants of 
the ${x}_+ x_-$ plane. The definitions \eqref{xpm-energy} have been chosen to 
describe the quadrant with $\pm x_\pm >0$. Fortunately this subtlety will not 
occur for 0A where it is necessary to consider both signs of $\epsilon$.

\subsection{Unitary Transformations and the Lax Formalism}

Consider some unitary transformation $\hat{U}$ acting on the system of 
fermions. We choose to employ a passive transformation picture where the state 
kets remain unchanged and the operators in (\ref{c=1algebra}) transform as 
$\hat{\mathcal{O}} \to \hat{U} \hat{\mathcal{O}} \hat{U}^{-1} $. In particular, 
we will label the transformed operators as
\bea
\hat{L}_{\pm} &=& \hat{U}\hat{x}_{\pm}\hat{U}^{-1}, \\
\hat{M} &=& \hat{U}\hat{\epsilon}\hat{U}^{-1}. \label{trans}
\eea
As our notation suggests, these transformed operators are precisely the Lax and 
Orlov-Shulman operators of the Toda lattice hierarchy when $\hat{U}$ is chosen 
appropriately. It follows trivially that the commutators and anti-commutators 
(\ref{c=1algebra}) are preserved by the unitary transformation under 
consideration. This leads to
\ben
\hat{L}_{\pm}\hat{L}_{\mp}=\hat{M}\pm\frac{i}{2}. \label{c1stringeqn}
\een
In the context of the Toda lattice hierarchy, \eqref{c1stringeqn} are known as 
the string equations. Unlike the usual treatment where separate $\hat{M}_\pm$ 
are introduced and then equated as an additional constraint, we posit only one 
Orlov-Shulman operator from the outset. Thus we are studying the constrained 
Toda hierarchy, \textit{ab initio}.

An integrable set of momentum mode deformations to the $c=1$ model are 
generated by the transformations
\begin{equation}
\hat{U}_{\pm}=e^{i \sum_{n>0} \tilde{b}_{\pm 
n}\left(\hat{\epsilon};\{t\}\right)\,(\pm \hat{x}_{\pm})^{-n/R}}e^{\mp 
i\tilde{\phi}\left(\hat{\epsilon};\{t\}\right)/2}e^{i\sum_{n>0} t_{\pm n} 
\,(\pm\hat{x}_{\pm})^{n/R}}.\label{eq:u-dress}
\end{equation}
The $t_{\pm n}$ are real constants\footnote{Although we will be more general in 
our analysis, in order to produce real deformations to the worldsheet action 
the signs of the $t$'s must be such that $t_n =-t_{-n}$ for all $n$. This will 
be true for the 0A theory as well.}, and the operators $\tilde{b}$ and 
$\tilde{\phi}$ are unspecified functions which vanish when all $t_{\pm n}=0$. 
We will constrain the undetermined functions $\tilde{b}$ and $\tilde{\phi}$ 
such that $\hat{U}_+ \equiv \hat{U}_- \equiv \hat{U}$. We understand 
$\hat{U}_+$ ($\hat{U}_-$) to be the natural form of $\hat{U}$ acting on the 
$\hat{x}_+$ ($\hat{x}_-$) operator. This is not to be confused with different 
bases; the expression $\hat{U}_+ = \hat{U}_-$ indicates an equality between 
operators, not that $\hat{U}_\pm$ are different representations of the same 
operator. Thus they are equal in any given basis.

The operators in (\ref{eq:u-dress}) are of interest since they allow us to 
represent momentum mode deformations to the worldsheet action in terms of 
transformations on the single fermion Hilbert space. The operators $(\pm 
\hat{x}_\pm)^{n/R}$ are the single-particle representations of the matrix model 
operators ${\rm Tr} (M \pm \dot{M})^{n/R}$ which create states of Euclidean 
momentum $\frac{n}{R}$ \cite{Jevicki:1993qn}. In the regions $x_\pm \to \infty$ 
the operators $\hat{U}_\pm$ behave like
\ben
\hat{U}_\pm \sim e^{\mp i\tilde{\phi}\left(\hat{\epsilon};\{t\}\right)/2} 
e^{i\sum_{n>0} t_{\pm n}\, \left(\pm \hat{x}_{\pm}\right)^{n/R}},
\een
demonstrating that asymptotically $\hat{U}_\pm$ creates a coherent state of 
tachyons (plus some zero mode). The factor with negative powers of 
$\hat{x}_\pm$ is present to allow the equality of $\hat{U}_+$ and 
$\hat{U}_-$.\com{ Thus the operator $\hat{U}=\hat{U}_\pm$ contains the same 
physical information as the generating functional of 
\cite{Dijkgraaf:1992hk}.\textbf{(Explain connection with generating functional 
better)}\com{Specifically, the generating functional is given by $ \mu^2 {\cal 
F} = \prod_{\epsilon<\mu} \langle \epsilon | \hat{U}_+ \hat{S} \hat{U}_-^{-1} 
|\epsilon \rangle $ where ${\cal F}$  and $\hat{S}$ are the generating 
functional and the single fermion scattering operator calculated in 
\cite{Dijkgraaf:1992hk}. }}

Now consider these operators in the energy basis. From (\ref{xpm-energy}), we 
have
\ben
\lt[e^{i\sum_{n>0}t_{\pm 
n}(\pm\hat{x}_{\pm})^{n/R}}\rt]_\epsilon=S^{\mp1/2}\left({\epsilon}\right)e^{i\sum_{n>0}t_{\pm 
n}\hat{\omega}^{\pm n/R}}S^{\pm1/2}\left({\epsilon}\right),
\een
and so find
\begin{equation}
\lt[\hat{U}_{\pm}\rt]_\epsilon =e^{i\sum_{n>0}b_{\pm 
n}\left({\epsilon};\{t\}\right)\hat{\omega}^{\mp n/R}}e^{\mp 
i\phi\left({\epsilon};\{t\}\right)/2}e^{i\sum_{n>0}t_{\pm n}\hat{\omega}^{\pm 
n/R}} S^{\pm 1/2} \left({\epsilon}\right),
\end{equation}
where we have introduced
\begin{eqnarray}
\phi\left({\epsilon};\{t\}\right) & \equiv & \tilde{\phi} 
\left({\epsilon};\{t\}\right) +\phi_{0}\left({\epsilon}\right),\nn\\
b_{\pm n} \left(\epsilon;\{t\}\right) & \equiv & \tilde{b}_{\pm n} 
\left({\epsilon};\{t\}\right) S^{\mp 1/2} \left(\epsilon\right) S^{\pm 1/2} 
\left({\epsilon}\mp in/R\right).
\end{eqnarray}

Using the energy basis and the equivalent forms of the transformation 
$\hat{U}$, we can make explicit the connection of the above with the Lax 
formalism. We write
\bea
\hat{L}_{\pm} & = & \hat{U}_{\pm}\hat{x}_{\pm}\hat{U}_{\pm}^{-1}, \nn\\
\hat{M}_\pm & = & \hat{U}_{\pm}{\hat\epsilon}\hat{U}_{\pm}^{-1}.
\eea
It should be emphasized that the above is simply a rewriting of \eqref{trans} 
and that
\ben
\hat{M} = \hat{M}_+  =  \hat{M}_- ,
\een
which follows from the equality of $\hat{U}$ and both $\hat{U}_\pm$. In the 
$\epsilon$ basis, we have
\bea
\lt[\hat{L}_{\pm}\rt]_\epsilon & = & \pm 
\hat{W}_{\pm}\hat{\omega}^{\pm1}\hat{W}_{\pm}^{-1}, \nn\\
\lt[\hat{M}_\pm\rt]_\epsilon & = & \hat{W}_{\pm}{\epsilon}\hat{W}_{\pm}^{-1}, 
\label{c1dress}
\eea
where $\hat{W}_{\pm}$ are the dressings
\ben
\hat{W}_{\pm} = e^{i\sum_{n>0}b_{\pm 
n}\left({\epsilon};\{t\}\right)\hat{\omega}^{\mp n/R}}e^{\mp 
i\phi\left({\epsilon};\{t\}\right)/2}e^{i\sum_{n>0}t_{\pm n}\hat{\omega}^{\pm 
n/R}}.
\een
We wish to emphasize that the $\hat{W}_\pm$, as functions of $\epsilon$ and 
$\ho$, are basis specific expressions.

In order to solve the string equation \eqref{c1stringeqn}, we first must have 
the expansion of the Orlov-Shulman operator in powers of the Lax operators. 
There are two equivalent expansions obtained by resumming the expansions of 
$\hat{M}_\pm$ in powers of $\ho$ in the energy representation in terms of the 
Lax operators
\ben
\hat{M}_\pm =  \hat{\epsilon} \mp \sum_{n>0} \frac{n t_{\pm 
n}}{R}\left(\pm\hat{L}_{\pm}\right)^{n/R}+\sum_{n>0}v_{\pm 
n}\left(\hat\epsilon;\{t\}\right) \left(\pm\hat{L}_{\pm}\right)^{-n/R}. 
\label{laxexpandM}
\een
The $v_{\pm n}(\hat\epsilon;\{t\})$ are undetermined functions which are in 
principle calculable. By introducing the functions $v_{\pm k}$ we have not 
introduced any more unknown functions, but rather simply reorganized the 
unknown $b_{\pm k}$ into the $v_{\pm k}$.

\subsection{Dispersionless Limit and Solving the String 
Equation}\label{c1displess}

The above formalism is a compact way to record an infinite hierarchy of 
finite-difference equations. These are obtained by expanding both sides of the 
string equation \eqref{c1stringeqn} in powers of $\ho$, with all factors of 
$\ho$ moved to the right. Matching the coefficients to each term in the series 
provides an infinite set of finite difference equations for the various 
undetermined functions in the operators $\hat{U}$. The most interesting of 
these functions is the zero-mode $\phi(\epsilon;\{t\})$ which provides the 
density of states via $\rho(\epsilon) =\frac{\partial_\epsilon \phi 
(\epsilon)}{2 \pi}$.

While the system of finite difference equations is in principle soluble, the 
study of these equations is not technically practical. Firstly, there are an 
infinite number of such equations and, secondly, even a finite set of finite 
difference equations is generally difficult to solve. Instead we will take 
advantage of the simplification occurring in the ``dispersionless limit'', when 
the lattice spacing goes to zero 
\cite{Alexandrov:2002fh,Alexandrov:2003ut,Kostov:2001wv,Kostov:2002tk}. In the 
fermion language, this is the classical limit, $\hbar \to 0$. Since we use 
$\hbar=1$ units this is accomplished by considering the regime $|\epsilon| \gg 
1$, which is the genus zero limit of the dual string theory. Since the 
Orlov-Shulman operator scales as $\epsilon$ and the Lax operators scale as 
$\sqrt{\epsilon}$, the string equation in this regime is
\ben
M = L_{-}L_{+}. \label{stringdispless}
\een
In this limit, the Lax operators are given by
\ben
L_{\pm}=\pm e^{-\partial_{\epsilon}\phi/2}\omega^{\pm1}\left(1+\sum_{k>0}a_{\pm 
k} \left(\epsilon;\{t\}\right)\omega^{\mp k/R}\right) +\ldots, \label{Lexpand}
\een
where the dots represent subleading terms in $\epsilon$ and the hat on $\omega$ 
as been dropped to indicate it is now a classical variable. The functions 
$a_{\pm k}(\epsilon;\{t\})$ are calculable from the functions in the 
transformation $\hat{U}$ but we will shortly solve for them using algebraic 
constraints, bypassing the need to ever know the precise form of the 
undetermined functions in $\hat{U}$.

We now take advantage of the expansion of $M$ in terms of the Lax variables 
\eqref{laxexpandM}. To simplify matters, we consider the case where all the 
couplings vanish except $t_{\pm n}$ for some particular $n$
\ben
M_{\pm}  =  \epsilon\mp\frac{n t_{\pm n}}{R}\lt(\pm 
L_{\pm}\rt)^{n/R}+\sum_{k>0}v_{\pm k}\left(\epsilon\right)\lt(\pm 
L_{\pm}\rt)^{-k/R}. \label{c1-Leq}
\een
These two expansions will effectively provide two string equations, $M_\pm = 
L_- L_+$, constraining the $a_{\pm k}$ coefficients strongly. Substituting 
\eqref{Lexpand} into the above, we find
\bea
M_\pm & = & \epsilon \mp \frac{nt_{\pm 
n}}{R}e^{-n\partial_{\epsilon}\phi/2R}\omega^{\pm n/R} \left(1+\sum_{k>0}a_{\pm 
k}\left(\epsilon\right)\omega^{\mp k/R} \right)^{n/R} \nn \\
&~ &  + \sum_{k>0}v_{\pm k}\left(\epsilon\right) e^{\frac{k}{R} 
\partial_\epsilon \phi} \omega^{\mp k/R} \lt(1+ \sum_{\ell>0}a_{\pm 
\ell}\left(\epsilon\right)\omega^{\mp \ell/R} \rt)^{-k/R}. \label{expandM}
\eea
Now compare \eqref{expandM} with
\ben
L_+ L_-  =  - 
e^{-\partial_{\epsilon}\phi}\left(1+\sum_{k>0}a_{+k}\left(\epsilon\right)\omega^{-k/R}\right) 
\left(1+\sum_{k>0}a_{-k}\left(\epsilon\right)\omega^{k/R}\right). 
\label{expandLL}
\een
Since the highest (lowest) power of $\omega$ in $M_\pm$ is $\omega^{\pm n/R}$, 
power matching in the string equation  implies $a_{\pm k}=0$ if $k>n$. 
Furthermore, upon examining other coefficients in the string equation one can 
see all the constraints are satisfied for $a_{\pm k}=0$ for $k\ne n$.

Bearing this in mind, we match the coefficients for the $\omega^{\pm n/R}$ 
terms in $M_\pm =L_+ L_-$ and obtain
\ben
a_{\mp n}\left(\epsilon\right)=\pm\frac{nt_{\pm 
n}}{R}e^{\left(1-n/2R\right)\partial_{\epsilon}\phi},
\een
which when substituted into the equation obtained from the $\omega^0$ terms 
results in\footnote{Although this result has been derived for $\epsilon<0$, it 
can be derived formally for positive energy as well.}
\begin{equation}
|\epsilon|-e^{-\partial_{\epsilon}\phi} = 
\frac{n^{2}}{R^{2}}t_{+n}t_{-n}e^{\left(1-n/R\right)\partial_{\epsilon}\phi}\left(\frac{n}{R}-1\right).
\label{eq:c=1MModes}
\end{equation}
We wish to emphasize the importance of \eqref{eq:c=1MModes}. This equation 
provides a transcendental equation for the density of states, and hence the 
free energy, in the presence of momentum mode perturbations. Our goal in the 
sections ahead is to obtain equations such as \eqref{eq:c=1MModes} for the 
density of states in 0A string theory perturbed by momentum modes.

As a quick consistency check of our method, we see that \eqref{eq:c=1MModes} 
produces the correct density of states for the $c=1$ theory when there is no 
perturbation,
\ben
\rho(\mu) = \left. \frac{1}{2\pi} \, \partial_\epsilon \phi 
\right|_{\epsilon=-\mu} = -\frac{1}{2\pi} \log \mu.
\een
For more details see, for example, \cite{Ginsparg:1993is}. A more non-trivial 
check is that \eqref{eq:c=1MModes}, with $n=1$ and a suitable redefinition of 
couplings, is precisely T-dual to the genus zero susceptibility equation 
obtained in \cite{Kazakov:2000pm} for the $c=1$ theory perturbed by winding 
modes.

The beauty of the procedure above is that there was never a need to know the 
undetermined functions $v_{\pm k}$. We have only assumed that they are such 
that it is possible to match the coefficients for the powers of $\omega$ which 
we do not examine. The fact that knowledge of the functions $v_{\pm k}$ is 
irrelevant to obtaining the equation for the density of states is not limited 
to the simple example above. Suppose that some finite number of couplings are 
turned on, but that $t_{\pm k}=0$ for $k>N$. Once again, a general argument 
indicates $a_{\pm k}$ vanish if $k>N$. Matching the coefficients for 
$\omega^{k/R}$ with $0 \le k \le N$ in $M_+ = L_+ L_-$, we obtain $N+1$ 
constraints, none of which contain $v_{\pm k}$. Similarly, examining the 
$\omega^{-k/R}$ coefficients with $0 < k \le N$ in $M_- = L_+ L_-$ leads to 
another $N$ equations which do not involve the $v_{\pm k}$. Together, these 
$2N+1$ constraints are sufficient to solve for $\partial_\epsilon \phi$ and the 
non-vanishing $a_{\pm k\le N}$. We will find in the next section that the most 
straightforward application of this method to 0A string theory will not be so 
tractable.

\section{Integrable Deformations of the 0A Matrix Model}

We would like to perform deformations of the 0A matrix model analogous
to those of the previous section. This has been previously attempted
\cite{Nakatsu:1995eg,Yin:2003iv,Park:2004yc,PandoZayas:2005tu}, but these 
authors
were only able to obtain solutions by setting $\mu$ or $q$ to zero. We will 
first demonstrate the intractability of the most straightforward organization 
of the integrable structure of the 0A matrix model. Then we consider an 
integrable structure based on the $2+1$ dimensional perspective for the 0A 
matrix model eigenvalues. Through this viewpoint we are able to find additional 
string equations which can be solved in the dispersionless limit with both 
$\mu$ and $q$ non-zero.

\subsection{The 0A Matrix Model}

It is well-known \cite{Douglas:2003up} that type 0A string theory in $d=2$ can 
be described by a gauged matrix model similar to that of the $d=2$ bosonic 
string. As is common with matrix models, the singlet sector can be reduced to 
the dynamics of a many-body problem of non-interacting fermions in $1+1$ 
dimensions. The single-particle Hamiltonian is
\ben
\hat{\epsilon}_{0A}=\frac{1}{2}\left(\hat{p}^{2}-\hat{x}^{2}+\frac{q^{2}-1/4}{\hat{x}^{2}}\right), 
\label{0AHam}
\een
where we consider $q$ to simply be some parameter. Proceeding as in the $c=1$ 
model, we put the operator algebra into the light-cone form 
\cite{Jevicki:1993qn}
\bea
\left[ \hat{B}_{+},\hat{B}_{-} \right] & = & 4i\hat{\epsilon}_{0A},\nn\\
\left[\hat{B}_{\pm},\hat{\epsilon}_{0A}\right] & = & \pm2i\hat{B}_{\pm},\nn\\
\left\{ \hat{B}_{+},\hat{B}_{-}\right\}  & = & 
2\left(\hat{\epsilon}_{0A}^{2}+q^{2}-1\right), \label{alg0A}
\eea
where
\ben
\hat{B}_{\pm} = \hat{x}_{\pm}^{2}+\left(q^{2}-1/4\right)\hat{x}^{-2}.
\een
We can easily obtain from \eqref{alg0A} the following identity
\ben
\hat{B}_\pm \hat{B}_\mp = \left( \hat{\epsilon}_{0A} \pm i \right)^2 +q^2.
\een
We will hereafter drop the subscript ``0A'' on the energy.

From the algebra \eqref{alg0A} we can find the energy basis representations
\ben
\left[ \hat{B}_\pm \right]_\epsilon = S^{\mp 1/2}(\epsilon,q) \ho^{\pm 2} 
S^{\pm 1/2} (\epsilon,q)\, , \label{0Ascatter}
\een
where $\ho$ is defined in \eqref{shift} and $S(\epsilon, q)$ is the scattering 
phase\footnote{The attentive reader may notice that this differs slightly from 
the result in \cite{Maldacena:2005he}. This is due to our somewhat different 
normalization of the fermion states. Note that the exclusion of the 
$i^{\left|q\right|}$ factor in \cite{Maldacena:2005he} does not affect the 
calculation of the density of states, $\rho(\epsilon) \sim \phi^\prime 
(\epsilon)$. See Appendix \ref{sub:0APhase} for more details.}
\begin{equation}
S\left(\epsilon,q\right)=2^{-i\epsilon}i^{\left|q\right|} 
\frac{\Gamma\left(1/2\left(1+\left|q\right|-i\epsilon\right)\right)}{\Gamma\left(1/2\left(1+\left|q\right|+i\epsilon\right)\right)} 
\equiv e^{i \phi_0 (\epsilon,q)}.
\label{eq:0A-S}
\end{equation}

\subsection{Intractability of the 0A Lax system}

Besides simplifying the 0A operator algebra the $\hat{B}_\pm$ play the role of 
momentum modes in the free fermion representation of 0A string theory 
\cite{Jevicki:1993qn}. Thus, following the analysis for the $c=1$ model, we 
wish to consider unitary transformations generated by powers of $\hat{B}_\pm$ 
to implement momentum mode perturbations. However, we will see that a 
straightforward application of this method will not lead to a solvable string 
equation in this instance.

For the 0A matrix model an integrable set of momentum mode deformations is 
generated by the unitary transformations
\ben
\hat{U}_{\pm}=e^{i\sum_{n>0}\tilde{b}_{\pm 
n}\left(\hat{\epsilon};q,\{t\}\right)\hat{B}_{\pm}^{-n/R}}e^{\mp 
i\tilde{\phi}\left(\hat{\epsilon};q,\{t\}\right)/2}e^{i\sum_{n>0} t_{\pm 
n}\hat{B}_{\pm}^{n/R}}.\label{0a-trans}
\een
Once again, $\tilde{\phi}$ and the $\tilde{b}$ are some undetermined functions 
which vanish when all of the couplings $\{t\}$, are turned off. As in the 
treatment of the $c=1$ theory, the operators \eqref{0a-trans} are constrained 
by $\hat{U}_+ = \hat{U}_-$. In the energy basis these operators are given by
\begin{equation}
\lt[\hat{U}_{\pm}\rt]_\epsilon =e^{i\sum_{n>0}b_{\pm 
n}\left({\epsilon};q,\{t\}\right)\hat{\omega}^{\mp 2n/R}}e^{\mp 
i\phi\left({\epsilon};q,\{t\}\right)/2}e^{i\sum_{n>0}t_{\pm n}\hat{\omega}^{\pm 
2n/R}} S^{\pm 1/2} \left({\epsilon},q\right), \label{U-energy}
\end{equation}
where we have introduced
\begin{eqnarray}
\phi\left({\epsilon};q,\{t\}\right) & \equiv & \tilde{\phi} 
\left({\epsilon};q,\{t\}\right) +\phi_{0}\left({\epsilon},q\right),\nn\\
b_{\pm n} \left(\epsilon;q,\{t\}\right) & \equiv & \tilde{b}_{\pm n} 
\left({\epsilon};q,\{t\}\right) S^{\mp 1/2} \left(\epsilon,q\right) S^{\pm 1/2} 
\left({\epsilon}\mp 2in/R,q\right).
\end{eqnarray}

We now define Lax and Orlov-Shulman operators as in the $c=1$ theory through 
unitary transformations on the 0A operators
\bea
\hat{L}_{\pm} &=& \hat{U}_\pm \hat{B}_{\pm}\hat{U}_\pm^{-1},\label{eq:L-B} 
\nn\\
\hat{M}_\pm &=& \hat{U}_\pm \hat{\epsilon}\hat{U}_\pm^{-1}. \label{0AOrlov}
\eea
where once again we have $\hat{M}_+ = \hat{M}_-$ due to the equivalence of the 
unitary transformations $\hat{U}_\pm$. In the energy basis these operators are 
represented as
\begin{eqnarray}
\lt[\hat{L}_{\pm}\rt]_\epsilon & = & 
\hat{W}_{\pm}\hat{\omega}^{\pm2}\hat{W}_{\pm}^{-1} ,\nn\\
\lt[{M}_\pm \rt]_\epsilon & = & \hat{W}_{\pm}{\epsilon}\hat{W}_{\pm}^{-1} ,
\end{eqnarray}
with the dressings given by
\ben
\hat{W}_{\pm} = e^{i\sum_{n>0}b_{\pm 
n}\left({\epsilon};q,\{t\}\right)\hat{\omega}^{\mp 2n/R}}e^{\mp 
i\phi\left({\epsilon};q,\{t\}\right)/2}e^{i\sum_{n>0}t_{\pm n}\hat{\omega}^{\pm 
2n/R}}.
\een
Since this is simply a unitary transformation, the operator algebra remains 
intact so we immediately get the string equation
\ben
\hat{L}_\pm \hat{L}_\mp = \left( \hat{M} \pm i \right)^2 +q^2 \, ,
\een
which was also obtained in \cite{Nakatsu:1995eg,Yin:2003iv}. This equation is 
perfectly correct but its use in reaching the density of states is limited. The 
problem lies in the appearance of a quadratic power of the Orlov-Shulman 
operator as we will see below.

In the dispersionless limit we take $\epsilon, q \gg 1$. The above string 
equation becomes
\ben
L_+ L_- = M^2 +q^2 \, ,\label{0Astringdispless}
\een
and the Lax operators can be expanded as
\ben
L_{\pm}= e^{-\partial_{\epsilon}\phi}\omega^{\pm2}\left(1+\sum_{k>0}a_{\pm k} 
\left(\epsilon;\{t\}\right)\omega^{\mp 2k/R}\right) \, .
\een
Once again, we examine a simple case where the only non-zero couplings are 
$t_{\pm n}$ for some $n$. The expansion of $M$ in terms of the Lax operators is 
similar to \eqref{c1-Leq}
\ben
M_{\pm}=\epsilon\mp\frac{2nt_{\pm n}}{R}L_{\pm}^{n/R}+\sum_{k>0}v_{\pm 
k}\left(\epsilon\right)L_{\pm}^{-k/R}.
\een

We examine the non-negative powers of $L_{\pm}$ that appear in $M_{\pm}^{2}$ 
(denoted $M_{\pm, \ge}^2$), since only these will contribute to the 
coefficients in which we are interested
\ben
M_{\pm,\ge}^{2}=\epsilon^{2}+\left(\frac{4nt_{\pm 
n}}{R}\right)^{2}L_{\pm}^{2n/R}\mp\frac{2nt_{\pm n}}{R}\,\left( \epsilon 
L_{\pm}^{n/R}+\sum_{k=1}^{n}L_{\pm}^{\left(n-k\right)/R}v_{\pm 
k}\left(\epsilon\right)\right). \label{Msqexpand}
\een
Compare this with
\ben
L_{+}L_{-}=e^{-\partial_{\epsilon}\phi}\left(1+\sum_{k>0}a_{+k}\left(\epsilon\right)\omega^{-2k/R}\right) 
\left(1+\sum_{k>0}a_{-k}\left(\epsilon\right)\omega^{2k/R}\right).
\een
Once again, matching of coefficients implies the $a_{\pm k}$ truncate. In this 
case we have $4n+1$ non-vanishing functions; $\partial_\epsilon \phi$ and the 
$a_{\pm k}$ for $k\le 2n$. However, one cannot obtain a closed set of equations 
for these variables. The $v_{\pm k}$ appear in the $2n+1$ equations one obtains 
from examining the coefficients to $\omega^{2k/R}$ with $0 \le k \le 2n$ in 
$L_+ L_- = M_+^2 +q^2$. This is also true of the $2n$ coefficients of 
$\omega^{-2k/R}$ with $0 < k \le 2n$ obtained from the $M_-$ string equation. 
Examining more coefficients will simply increase the number of $v_{\pm k}$ 
involved. One must then solve the whole infinite set of equations.

One can see that the origin of the above intractability is the appearance of 
the unknown functions $v_{\pm k}$ multiplying non-negative powers of $L_\pm$ in 
the string equation once \eqref{Msqexpand} is substituted into 
\eqref{0Astringdispless}. This would be alleviated were the string equation 
linear in the Orlov-Shulman operator. We now turn to such a way of organizing 
the 0A system.

\subsection{Flux and the Complex Basis}
We have found that the Toda Lattice based on shifts of the single-particle 
energy do not lead to solvable string equations. In fact, it appears as if we 
are missing information. To remedy this, we recall that the one-dimensional 
Hamiltonian \eqref{0AHam} can be understood as the effective radial Hamiltonian 
for fermions moving in a two-dimensional harmonic oscillator 
\cite{Douglas:2003up}. In this description, the RR-flux is the angular momentum 
operator, $\hat{q}$. In terms of the Cartesian coordinates and momenta in the 
plane, the energy and angular momentum are given by
\bea
\hat{\epsilon} & = & 
\frac{1}{2}\left(\hat{p}_{1}^{2}+\hat{p}_{2}^{2}-\hat{x}_{1}^{2}-\hat{x}_{2}^{2}\right) 
\, ,  \\
\hat{q} & = & \hat{x}_1 \hat{p}_2 - \hat{x}_2 \hat{p}_1 \, .
\eea
It is convenient to complexify these operators
\begin{eqnarray}
\hat{z}_{\pm} & \equiv & \left(\frac{\hat{p}_{1} \pm 
\hat{x}_1}{\sqrt{2}}\right) +i\left(\frac{\hat{p}_{2} \pm 
\hat{x}_2}{\sqrt{2}}\right),\\
\hat{y} & \equiv & \hat{\epsilon}+i\hat{q} \, , \label{complex}
\end{eqnarray}
which obey the algebra
\bea
\left[\hat{\bar{z}}_{+},\hat{z}_{-}\right] & = &2i \, , \nn \\
\left[\hat{z}_{-},\hat{y}\right] & = &-2i \hat{z}_{-} \, ,\nn \\
\left[\hat{\bar{z}}_{+},\hat{y}\right] & = & 2i \hat{\bar{z}}_{+} \, 
,\label{complexsp4}
\eea
where the bar above an operator denotes its Hermitian conjugate. Other 
commutators are trivial or obtained by Hermitian conjugation of 
(\ref{complexsp4}). For convenience we also record the relation
\ben
\hat{z}_{-}\hat{\bar{z}}_{+}  =  \hat{y}-i.\label{0Aundressed}\\
\een

We can obtain the previous algebra (\ref{alg0A}) by noting that
\ben
\hat{B}_\pm = \hat{z}_\pm \hat{\bar{z}}_\pm \, .\label{Bz}
\een
Recalling that $\hat{B}_\pm$ are the vertex operators for momentum modes, it is 
evident that in the complex coordinates we should consider unitary 
transformations generated by powers of $\hat{z}_\pm \hat{\bar{z}}_\pm$ to 
implement momentum mode deformations. This is done in the following section. 
But first let us obtain the $(\epsilon,q)$-basis for the operators above.

Define the shift derivative $\hat{\eta} = e^{2 i \partial}$ with Hermitian 
conjugate $\hat{\bar\eta} = e^{2 i \bar\partial}$ where
\bea
\partial & \equiv & \partial_y = 
\frac{1}{2}\left(\partial_{\epsilon}-i\partial_{q}\right), \nn \\
\bar{\partial} & \equiv & \partial_{\bar{y}} = 
\frac{1}{2}\left(\partial_{\epsilon}+i\partial_{q}\right).\label{eq:CompDeriv}
\eea
Examining the algebra (\ref{complexsp4}), we see that $\hat{\bar{z}}_+ \sim 
\hat{\eta}$ and $\hat{z}_- \sim \hat{\eta}^{-1}$ will reproduce the commutators 
with $\hat{y}$. To reproduce the final commutator and \eqref{0Aundressed}, we 
dress the appropriate power of $\hat\eta$ with the scattering 
phase\footnote{Without loss of generality we have assumed that $q>0$. For 
$q<0$, it is important that $|q|$ appears in the scattering amplitude dressing 
$\hat{\eta}$. However, this eventually leads to the same susceptibility 
equation \eqref{0asol:cplx}.}
\bea
\left[\hat{z}_{-}\right]_{y,\bar{y}} & = & S^{1/2}\left(y,\bar{y}\right) 
\hat{\eta}^{-1}S^{-1/2}\left(y,\bar{y}\right), \nn \\
\left[\hat{\bar{z}}_{+}\right]_{y,\bar{y}} & = & S^{-1/2}\left(y,\bar{y}\right) 
\hat{\eta} \, S^{1/2}\left(y,\bar{y}\right) \, ,\label{0Aeqbasis}
\eea
with other operators defined by Hermitian conjugation and $S\lt(y,\bar{y}\rt)$ 
given by
\ben
S\left(y,\bar{y}\right) = 2^{-i \, Re(y)} i^{Im(y)} \frac{\Gamma\lt( 
\half\lt(1-iy\rt)\rt)}{\Gamma\lt( \half\lt(1+i\bar{y}\rt)\rt)}.
\een

\subsection{Complexified Lax Formalism}

Now we act with some unitary operator, $\hat{U}$, obtaining the complex Lax and 
Orlov-Shulman operators\footnote{Note that the algebra of all operators 
quadratic in the $z$'s is $\mathfrak{sp}_4$.  It may be possible to construct 
generalized Toda lattices from other Lie algebras, building in and out Lax 
operators from the positive and negative roots, and Orlov-Shulman operators 
from the elements of the Cartan subalgebra. Such a construction is a subject 
for future research. }
\begin{eqnarray}
\hat{Z}_{\pm} & \equiv & \hat{U}\hat{z}_\pm\,\hat{U}^{-1},\\
\hat{Y} & \equiv &\hat{U}\hat{y}\,\hat{U}^{-1},
\end{eqnarray}
and similarly for their Hermitian conjugates. The algebra \eqref{complexsp4} is 
preserved by the unitary transformation as is, most importantly, the string 
equation
\ben
\hat{Z}_{-}\hat{\bar{Z}}_{+}  =  \hat{Y}-i. \label{0Astring}
\een
This is essentially a complexified version of the string equation for the $c=1$ 
theory. To implement momentum mode perturbations we use the unitary 
transformations \eqref{0a-trans} generated by $B_\pm$. The $y$ basis 
representations are easily obtained from \eqref{U-energy} by noting that $\ho^2 
= \hat{\eta}\hat{\bar\eta}$
\begin{equation}
\lt[\hat{U}_{\pm}\rt]_{y,\bar{y}} =e^{i\sum_{n>0}b_{\pm 
n}\left(y,\bar{y};\{t\}\right)\left(\hat{\eta} \hat{\bar\eta}\right)^{\mp 
n/R}}e^{\mp i\phi\left(y,\bar{y};\{t\}\right)/2}e^{i\sum_{n>0}t_{\pm 
n}\left(\hat{\eta} \hat{\bar\eta}\right)^{\pm n/R}} S^{\pm 1/2} 
\left(y,\bar{y}\right).
\end{equation}
As with the previous examples we impose $\hat{U}_+ = \hat{U}_-$ and define the 
transformed operators
\bea
\hat{Z}_{\pm} & \equiv & \hat{U}_\pm \hat{z}_\pm\,\hat{U}_\pm^{-1},\\
\hat{Y}_\pm & \equiv &\hat{U}_\pm \hat{y}\,\hat{U}_\pm^{-1} \, ,
\eea
where $\hat{Y}_+ =\hat{Y}_- = \hat{Y}$. The $y$ basis expressions are given by
\bea
\lt[ \hat{Z}_{-}\rt]_{y,\bar{y}} & = & \hat{W}_- \hat{\eta}^{-1} 
\,\hat{W}_-^{-1},\\
\lt[ \hat{\bar{Z}}_{+}\rt]_{y,\bar{y}} & = & \hat{W}_+ \hat{\eta} 
\,\hat{W}_+^{-1},\\
\lt[ \hat{Y}_\pm\rt]_{y,\bar{y}} & = &\hat{W}_\pm \hat{y}\,\hat{W}_\pm^{-1} ,
\eea
where the dressings are
\ben
\hat{W}_{\pm} =e^{i\sum_{n>0}b_{\pm 
n}\left(y,\bar{y};\{t\}\right)\left(\hat{\eta} \hat{\bar\eta}\right)^{\mp 
n/R}}e^{\mp i\phi\left(y,\bar{y};\{t\}\right)/2}e^{i\sum_{n>0}t_{\pm 
n}\left(\hat{\eta} \hat{\bar\eta}\right)^{\pm n/R}}.
\een
By reorganizing the $\hat\eta$ expansions of the two equivalent forms of 
$\hat{Y}$ we find the expansions of the Orlov-Shulman operator in terms of the 
Lax operators
\ben
\hat{Y}_{\pm} =  \hat{y} \mp \sum_{k >0} \frac{2k}{R}t_{\pm k} 
\hat{Z}_{\pm}^{\,k/R}\hat{\bar{Z}}_{\pm}^{\,k/R}+ \sum_{k>0}v_{\pm 
k}\left(\hat{y},\hat{\bar{y}}\right)\hat{Z}_{\pm}^{\,-k/R}\hat{\bar{Z}}_{\pm}^{\,-k/R} 
\, . \label{YexpandZ}
\een
In the dispersionless limit, we simply drop the hats in the above. The string 
equation becomes
\ben
Y_\pm = Z_- \bar{Z}_+ \, ,\label{complexdispless}
\een
and the Lax variables have expansions
\bea
Z_- & = & e^{-\partial \phi} \eta^{-1} \left(1+ \sum_{k>0} a_{- k} (y,\bar{y}) 
\lt(\eta\bar\eta\rt)^{k/R}\right) , \\
\bar{Z}_+ & = & e^{-\partial \phi} \eta \left(1+ \sum_{k>0} {a}_{+ k} 
(y,\bar{y}) \lt(\eta\bar\eta\rt)^{-k/R}\right) ,
\eea
and similarly for their conjugates. Conveniently and unsurprisingly, we obtain 
only powers of $\eta\bar{\eta} =\omega^2$ in the expansion of 
\eqref{complexdispless}.

To be explicit, assume all $t_{\pm k}=0$ except $t_{\pm n}$. The matching of 
coefficients once again implies only $a_{\pm n}$ are non-vanishing. Since the 
$v_{\pm k}$ do not multiply positive powers of the Lax variables in the string 
equation, we have few enough variables to solve the equations imposed by 
matching of coefficients. We obtain
\ben
a_{\pm n} = \pm \frac{2n}{R} t_{\mp n} e^{2 \partial\phi} e^{-\frac{n}{R} 
\lt(\partial \phi + \bar{\partial} \phi \rt)} \, ,
\een
by matching the $\left(\eta\bar\eta\right)^{\pm n/R}$ coefficients. The order 
$\left(\eta\bar\eta\right)^0$ coefficient matching then gives the complex 
susceptibility equation
\ben
e^{-2\partial \phi} -y = \frac{4 n^2}{R^2} t_{+n} t_{-n} e^{-\frac{2n}{R} 
\lt(\partial \phi + \bar{\partial} \phi \rt)} 
\left(\left(1-\frac{n}{R}\right)e^{2\partial\phi} -\frac{n}{R} 
e^{2\bar{\partial} \phi}\right). \label{0asol:cplx}
\een
The above equation is perhaps more transparent when split into real and 
imaginary parts
\bea
\epsilon-e^{-\partial_{\epsilon}\phi}\cos\left(\partial_{q}\phi\right) & = & 
\frac{4 n^2 
t_{-n}t_{+n}}{R^{2}}e^{\left(1-2n/R\right)\partial_{\epsilon}\phi}\left(\frac{2n}{R}-1\right) 
\cos\left(\partial_{q}\phi\right), \nn\\
q-e^{-\partial_{\epsilon}\phi}\sin\left(\partial_{q}\phi\right) & = & \frac{4 
n^2 
t_{-n}t_{+n}}{R^{2}}e^{\left(1-2n/R\right)\partial_{\epsilon}\phi}\sin\left(\partial_{q}\phi\right).\label{0asol}
\eea

The equation \eqref{0asol:cplx} (or alternatively the real and imaginary parts 
\eqref{0asol}) constitutes the main result of our paper, as it contains the 
information necessary to extract the genus zero susceptibility for 0A string 
theory in the presence of momentum mode perturbations. A few comments are in 
order. First, we would like to emphasize that \eqref{0asol:cplx} has been 
derived in full generality, with both non-vanishing cosmological constant $\mu$ 
and RR flux $q$. To the best of our knowledge, this is the first instance in 
which a set of string equations have been solved in a flux background. Second, 
we have arrived at these results by exploiting the integrability of the 0A 
matrix model. We have found that perturbations to the matrix model can be cast 
as a complexified Toda system, and have derived the string equations 
constraining it. In the dispersionless limit, these string equations yield 
\eqref{0asol:cplx}.

\section{Analysis of the 0A Susceptibility Equations}

We wish to analyze the complex susceptibility equation \eqref{0asol:cplx}, with 
the goal of obtaining the phase diagram of the system. Given the transcendental 
nature of the equation, this is a difficult task. We begin by specializing to 
the case of no Ramond-Ramond flux and perform an analysis T-dual to that of 
\cite{Kazakov:2000pm}. We find critical behavior which obstructs smoothly 
moving through parameter space to the momentum mode condensate phase for 
$R>2n$. We then proceed to the case of non-trivial flux and explore the 
critical behavior present there.

Before diving into the intricacies of these transcendental equations, let us 
digress briefly to make some general comments. Firstly, the perturbation 
couplings $t_{\pm n}$ must be such that the worldsheet Lagrangian is deformed 
by a real operator. For the 0A theory, the unitary transformations generating 
the momentum perturbations have the asymptotic forms $\hat{U}_\pm \sim e^{i 
t_{\pm n} \hat{B}_\pm^{n/R}}$, which act on the single-fermion Hamiltonian as 
$\hat{H} \to \hat{H} \mp \frac{2 n t_{\pm n}}{R} \hat{B}_{\pm}^{n/R}$. This 
translates, in the absence of flux, to deforming the worldsheet Lagrangian as
\ben
{\cal L} \to {\cal L} - t_{n} {\cal V}_{2n/R} + t_{-n} {\cal V}_{-2 n/R}. 
\label{lagdef}
\een
Given the form of the vertex operators ${\cal V}_{p} \sim e^{ipX}$, we see that 
a real deformation requires that $t_n = - t_{-n} \equiv t$ for some real $t$. 
This is not expected to change in the presence of RR flux. We will thus make 
the substitution $t^2 = -t_n t_{-n}$ hereafter.

Secondly, we will pursue a thermodynamic analysis of 0A string theory in what 
follows. This can be accomplished with our susceptibility equations since 
$\partial_\epsilon \phi (\epsilon)= 2\pi \rho(\epsilon)$, where 
$\rho(\epsilon)$ is the density of states. The grand canonical partition 
function, with fixed $q$, is then
\ben
\ln {\cal Z}=\frac 1{2\pi}\int_{-\infty}^\infty d\epsilon \, \partial_\epsilon
\phi(\epsilon)\ln(1+e^{2\pi R(\mu - \epsilon)}).
\een
It follows that
\bea
\partial_\mu \ln {\cal Z} & = &R \int_{-\infty}^\infty d\epsilon \, 
\frac{\partial_\epsilon
\phi(\epsilon)}{1+e^{2\pi R(\epsilon-\mu)}} \nn \\
        & \approx & R \int_{-\infty}^\mu d\epsilon \, \partial_\epsilon 
\phi(\epsilon) \, ,
\eea
where in the second line, we have kept only the leading term as $|\mu| \to 
\infty$. We thus arrive at the genus zero expression for the susceptibility
\ben
\chi(\mu) \equiv \frac{1}{R}\partial_\mu^2 \ln {\cal Z} = \partial_\mu \phi 
(\mu). \label{suscept_def}
\een
In keeping with this notation we will make the replacement $\epsilon =\mu$ 
henceforth.

\subsection{Relevant Perturbations at $q=0$}\label{relevant}

We are studying in this paper deformations of the string worldsheet theory by 
tachyon vertex operators of definite Euclidean momentum. In order to have an 
effect on the long wavelength dynamics of the theory, and hence the 
semi-classical spacetime geometry, this must be a relevant deformation in the 
sense of worldsheet renormalization group flow. This consideration puts a bound 
on the radius of compactification that we consider.

Let us consider the bosonic string for illustrative purposes. Before gauge 
fixing the worldsheet metric, the vertex operators we are considering are 
simply
\ben
{\cal V}_{p}=e^{ipX}, \label{vert}
\een
and have conformal dimension $\Delta = \frac{p^2}{4}$ in $\alpha^\prime = 1$ 
units. If ${\cal V}_{p}=e^{ipX}$ is to be relevant then $|p| < 2$. This same 
bound applies to the canonical ghost picture\footnote{That is, picture 
$(-1,-1)$ for NS-NS operators and picture $(-\half,-\half)$ for R-R operators.} 
Type 0 vertex operators at zero Ramond-Ramond flux in $\alpha^\prime =\half$ 
units. To obtain the susceptibility equation \eqref{0asol:cplx} for the 0A 
matrix model, we have considered perturbations with momenta $p = \pm 
\frac{2n}{R}$. Thus, for a given $n$, we have a bound on the Euclidean time 
radius, $R>n$. Although this has been argued for the case of vanishing flux we 
will shortly see indications that this bound should be extended to $q \ne 0$.

\subsection{Scaling Behavior of $c=1$}\label{scaling}


We begin by considering the scaling behavior in $\mu$ of the perturbed 0A 
theory with $q=0$. One might expect perturbed 0A with no flux to be identical 
to the perturbed $c=1$ matrix model, and in this section we will highlight the 
similarities of these two theories. In section \ref{sec:CritSurf-qneq0} we will 
see that 0A contains novel behavior resulting from divergences in the 
$q$-derivatives of the free energy, even at strictly vanishing flux.

To explore the similarities between $c=1$ and 0A with $q=0$, it is most useful 
to use the component form of the susceptibility equation. For vanishing flux, 
the second equation in \eqref{0asol} is satisfied with $\sin \lt(\partial_q 
\phi\rt)=0$. When all of the $t_{\pm n}$ vanish, $\cos \lt(\partial_q \phi \rt) 
= {\rm sign}(\mu)$ at large $|\mu|$ and $q=0$; we will assume this is so when 
the perturbations are non-trivial. Thus the first equation of \eqref{0asol} 
reduces to
\ben
|\mu|=e^{-\partial_\mu \phi}+\frac{4n^2t^2}{R^2}
\bigg(1-\frac {2n}R\bigg)e^{(1-2n/R)\partial_\mu \phi}. \label{c=1sol}
\een
This is identical to the result \eqref{eq:c=1MModes} for the $c=1$ theory 
perturbed by modes with momentum $p=\pm \frac{2n}{R}$. This factor of 2 is 
accounted for by the fact that only perturbations by $\hat{B}_\pm \sim 
\left(\hat{x}_\pm\right)^2$ have been used in our treatment of the 0A theory. 
Our result \eqref{c=1sol} is thus T-dual to that obtained through studies of 
the $c=1$ theory perturbed by even winding modes 
\cite{Kazakov:2000pm,Kostov:2001wv}.

We find that \eqref{c=1sol} can be expressed more simply by using the 
positive-definite dimensionless parameter\footnote{The parameter $\Lambda$ is a 
natural variable since, being linear in the coupling $t^2$, it organizes the 
perturbative expansion around the 0A string without flux. Another natural 
variable, $\tilde\Lambda =\Lambda^{-\frac{1}{2(1-n/R)}}$, is linear in $|\mu|$ 
and associated with the expansion near the theory with momentum mode 
condensate.}
\ben
\Lambda = \frac{4 n^2}{R^2} t^2 |\mu|^{\frac{2n}{R} -2}. \label{lambda}
\een
This variable replaces the perturbation coupling $t^2$ with a coupling that 
runs with the worldsheet cosmological constant. Recalling that $R>n$ for any 
relevant perturbation, we see that the unperturbed theory is located at 
$\Lambda=0$ by sending $t^2 \to 0$ and $|\mu| \to \infty$. Similarly, 
condensation of the momentum mode perturbation occurs as $\Lambda \to \infty$ 
via the opposite limits, $t^2 \to \infty$ and $|\mu| \to 0$. For $R<n$ these 
limits break down, and the variable $\Lambda$ no longer has a clear limit in 
which the theory is unperturbed. Although there will not be a bound from 
worldsheet relevance when we study $q \ne 0$, we will see a similar behavior in 
the corresponding dimensionless variable.

To take advantage of the dimensionless variable we introduce the reduced 
susceptibility $\tilde{\chi}$ through
\ben
\chi \equiv \partial_\mu \phi = \tilde{\chi} - \frac{1}{2(1-\frac{n}{R})} \log 
\lt( \frac{4 n^2}{R^2} t^2 \rt).
\label{suscept_red}
\een
Then \eqref{c=1sol} reduces to
\ben
\Lambda^{\frac{1}{2(n/R-1)}}=e^{-\tilde\chi}+\lt(1-\frac{2n}R\rt) 
e^{(1-2n/R)\tilde\chi},
\label{c1reduced}
\een
indicating that $\tilde \chi$ is function only of $\Lambda$ and $R$.

\subsection{Critical Behavior of  $c=1$} \label{critq0}

\begin{figure}[ht]
\begin{center}
\leavevmode
\epsfxsize 11cm \epsfbox{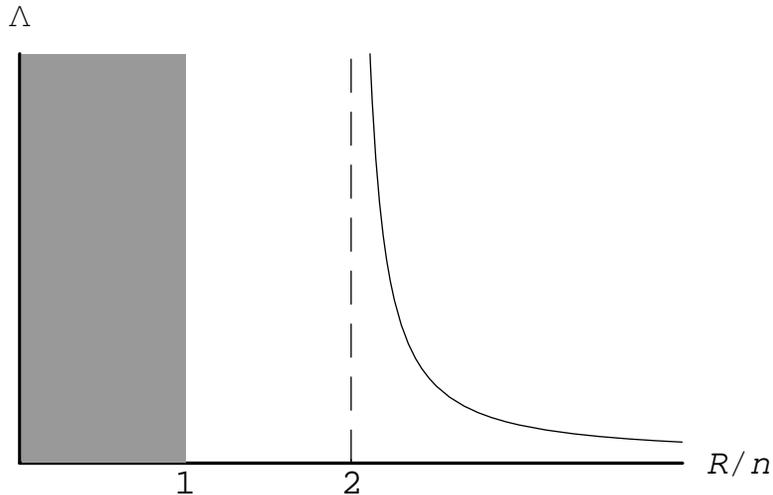}
\end{center}
\caption{\footnotesize The phase diagram for $c=1$ theory perturbed by momentum 
modes. For
$\Lambda=0$, the system is in the unperturbed 0A phase, and for
$\Lambda\to\infty$, in the momentum mode condensate phase.
For $n<R<2n$
one can transition freely between these two phases, whereas for $R>2n$, there
will be an obstruction given by the curve $\Lambda_c (R)$, associated with the 
phase transition to $c=0$. For $\Lambda > \Lambda_c(R)$, there are no solutions 
to the reduced susceptibility equation \eqref{c1reduced}. The shaded region for 
$R<n$ is forbidden due to the irrelevance of the operators ${\cal V}_{\pm 
n/R}$.}
    \label{Plot1}
\end{figure}

The critical behavior of $c=1$ string theory can be inferred from 
\eqref{c1reduced}. We differentiate this relation with respect to $\Lambda$ and 
obtain
\bea
\partial_\Lambda \tilde \chi  & = & \frac{\Lambda^{\frac{1}{2(n/R-1)}-1} 
e^{\tilde\chi}}{2(n/R-1)} \left[ \lt(1-\frac{2n}{R}\rt)^2 
e^{2(1-n/R)\tilde\chi} -1\rt]^{-1} \label{divdchi1} \\
     & = & \frac{\Lambda^{\frac{1}{2(1-n/R)}-1} e^{\tilde\chi}}{2(n/R-1)} \left[ 
\lt(1-\frac{2n}{R}\rt) \left(2-\frac{2n}{R}\right) e^{ (1-2n/R) \tilde\chi} - 
\Lambda^{\frac{1}{2(n/R-1)}} \rt]^{-1} \, . \label{divdchi2}
\eea
In the second line we have substituted the susceptibility equation into 
\eqref{divdchi1} to obtain a relation for later comparison with the case $q \ne 
0$. As is clear from \eqref{divdchi1}, $\partial_\Lambda \tilde \chi$ has a 
pole at a finite value of $\tilde \chi$
\ben
\tilde \chi_c = \frac{1}{\lt(n/R-1\rt)} \log \lt|1 -\frac{2n}{R}\rt| 
\label{pole}.
\een
To confirm that this is critical behavior, we must check that the equation of 
state is satisfied. Substituting \eqref{pole} into \eqref{c1reduced}, we obtain 
the critical surface in the $\Lambda-R$ plane
\ben
\Lambda_c^{\frac{1}{2(n/R-1)}} = {\rm sign} \left(1-\frac{2n}{R}\right) 
\lt(2-\frac{2n}{R}\rt) 
\lt|1-\frac{2n}{R}\rt|^{\frac{n}{R}\lt(1-\frac{n}{R}\rt)^{-1}}. \label{lc_imp}
\een
Since $\Lambda$ is positive definite, we see that there is no critical behavior 
for $n<R<2n$; this can also be discerned from \eqref{divdchi2}. We discard the 
solution for $R<n$ as the perturbation is irrelevant in that region and solve 
\eqref{lc_imp} for $\Lambda_c$
\ben
\label{ceq1critsurf}
\Lambda_c(R)=\bigg(1-\frac{2n}R\bigg)^{-2n/R}\bigg[2\bigg(1-\frac nR\bigg)
\bigg]^{-2(1-n/R)}, \quad\quad R>2n \, .
\een
This cleanly divides the parameter space of the theory as can be seen in Figure 
(\ref{Plot1}).
Thus, this critical behavior provides an obstruction to smoothly perturbing 
from the unperturbed 0A phase to the momentum mode condensate phase with 
$R>2n$. Beyond this, for $\Lambda > \Lambda_c(R)$, there do not exist real 
solutions to the equation of state \eqref{c1reduced}.

\begin{figure}[ht]
\begin{center}
\leavevmode
\epsfxsize 7cm \epsfbox{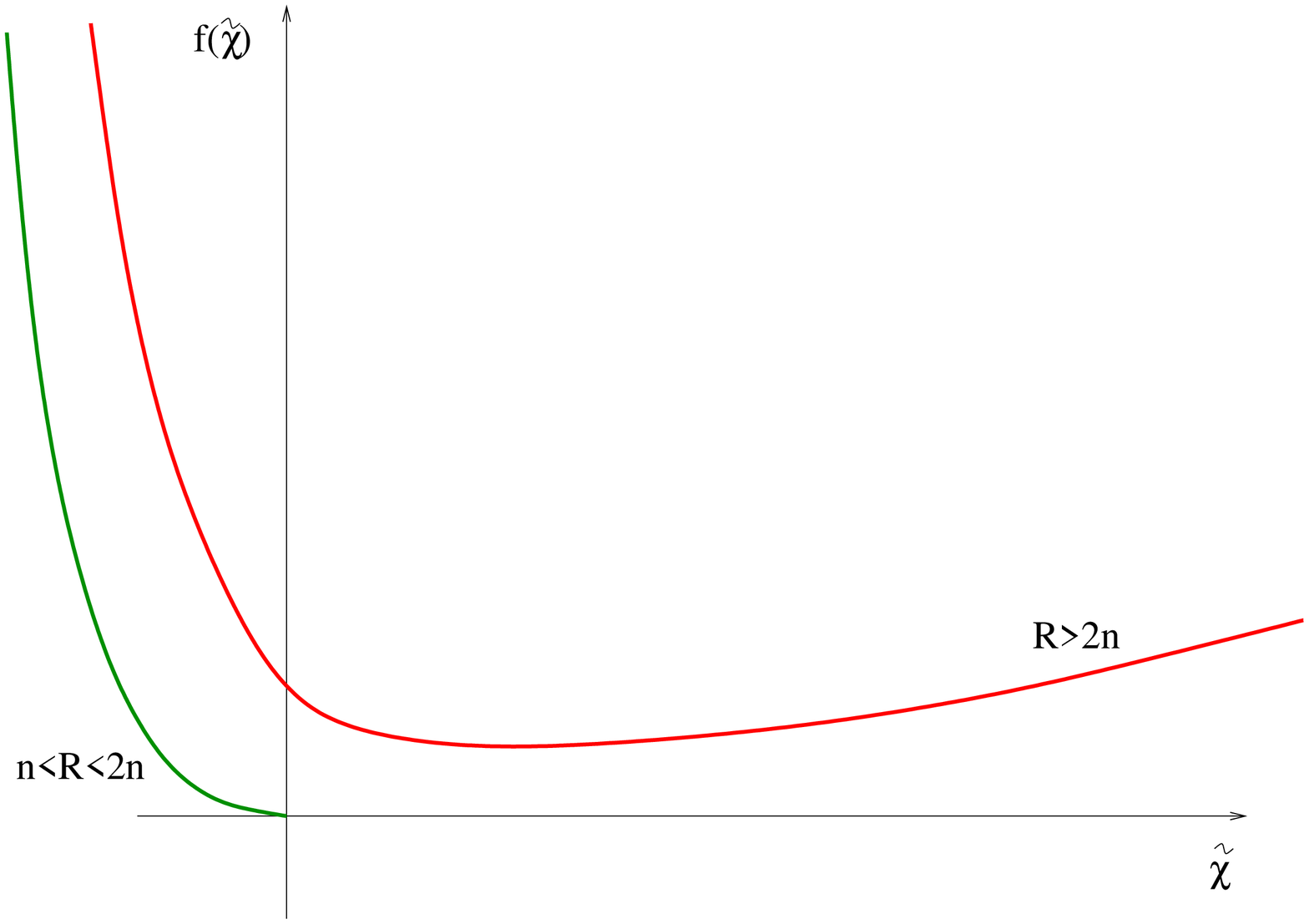}
\end{center}
\caption{\footnotesize  The right-hand-side of \eqref{c1reduced} is plotted
as a function of $\tilde\chi$ for two values of R, as indicated.   }
    \label{Plot2}
\end{figure}

Alternatively, let us plot the right-hand-side of \eqref{c1reduced} (call it
$f(\tilde\chi)$) in Figure (\ref{Plot2}). The behavior described above is 
associated with a local minimum at $\tilde\chi=\tilde\chi_c$, where 
$f(\tilde\chi_c) = \Lambda_c^{\frac{1}{2(n/R-1)}}$. Again, we find that there 
exist no solutions for $\Lambda > \Lambda_c(R)$.
More generally, all local extrema of $f(\tilde\chi)$ will be associated with 
critical
points since $\partial_\Lambda \tilde\chi\sim 1/(\partial_{\tilde\chi}f)$.

The above critical behavior can be understood thermodynamically by recalling 
from \eqref{suscept_def} that $\chi = \frac{1}{R} \partial_\mu^2 \ln {\cal Z}$. 
Then, using \eqref{suscept_red}, we see that $\Lambda_c(R)$ is the surface 
where there is a divergence in
\ben
\lt. \partial_\mu^3 \ln {\cal Z} \, \rt|_{t,R} \sim \lt. \partial_\Lambda 
\tilde\chi \,  \rt|_{R}.
\een
In the vicinity of this critical surface one has the behavior
\ben
\partial_\mu^3 \ln Z \sim (\Lambda-\Lambda_c)^{-1/2},
\label{c=0}
\een
typical of pure $d=2$ gravity, known sometimes as a $c=0$ model 
\cite{Hsu:1992cm}. The physical picture of this phase transition is that the 
worldsheet field, $X$, associated with Euclidean time freezes out by settling 
into one of the minima of the cosine potential in \eqref{lagdef}.

\subsection{Perturbative Analysis with Non-trivial Flux}\label{pert_check}

Having analyzed the susceptibility equations at $q=0$, we would now like to 
study them in the presence of RR flux.  To gain some confidence in the 
veracity of \eqref{0asol:cplx}, we perform two consistency checks.

We first examine the limit of vanishing momentum mode perturbations; this is 
the case studied in \cite{Maldacena:2005he}.  Solving  \eqref{0asol:cplx} with 
$t^2=0$ we find
\ben
\partial_\mu \phi(\mu) =  (\partial+\bar\partial)\phi=-\frac{1}{2}\ln(y\bar 
y)=-\frac{1}{2}\ln\lt(\mu^2 +q^2\rt),\label{0aleadingterm}
\een
which does indeed agree with the known result to leading order in $\mu^2+q^2$. 
We shall use this expression for the unperturbed partition function as a 
boundary condition for the non-linear PDE system \eqref{0asol}.

We next utilize perturbation theory to expand about the solution 
\eqref{0aleadingterm}. Through the 0A string/matrix model duality we expect 
that the worldsheet theory perturbed as in \eqref{lagdef} should be related to 
the perturbed matrix model free energy through
\begin{equation}
\ln\mathcal{Z}=\left\langle 
e^{t_{+n}\mathcal{T}_{2n/R}-t_{-n}\mathcal{T}_{-2n/R}}\right\rangle .
\end{equation}
In perturbation theory, the right-hand side is evaluated as a sum of tachyon 
correlators in the unperturbed worldsheet theory. Through \eqref{suscept_def}, 
we can then interpret $\partial_{\mu}\phi$ as a generating
functional for the tachyon correlators on the sphere
\begin{equation}
\partial_{\mu}\phi=R^{-1}\partial_{\mu}^{2}\ln\mathcal{Z}=-\frac{1}{2}\ln\left(\mu^2+q^2\right)+R^{-1}\sum_{m>0}\frac{t^{2m}}{\left(m!\right)^{2}}\partial_{\mu}^{2}\left\langle 
\mathcal{T}_{2n/R}^{m}\mathcal{T}_{-2n/R}^{m}\right\rangle . 
\label{dphi_generate}
\end{equation}
In order to compute these correlators, we first eliminate $\partial_{q}\phi$
from \eqref{0asol},
\begin{equation}
e^{-2\partial_{\mu}\phi}=\frac{\mu^{2}}{\left[1-\left(2n/R-1\right)\left(4n^{2}t^{2}/R^{2}\right)e^{-2\left(n/R-1\right)\partial_{\mu}\phi}\right]^{2}}+\frac{q^{2}}{\left[1-\left(4n^{2}t^{2}/R^{2}\right)e^{-2\left(n/R-1\right)\partial_{\mu}\phi}\right]^{2}},\label{0aradeq}
\end{equation}
then expand $\partial_{\mu}\phi$ in powers of $t^{2}$ around 
\eqref{0aleadingterm}. In
appendix \ref{app:correlators} we compute these tachyon correlators
to $\mathcal{O}\left(t^{6}\right)$ and find that they agree with those 
calculated
in \cite{PandoZayas:2005tu}. This gives us great confidence in our result for 
general flux and momentum mode perturbation.

\subsection{Critical behavior in Flux Vacua}
\label{sec:CritSurf-qneq0}

We are now in position to construct
the $d=2$ 0A phase diagram. We are going to follow the same
strategy as outlined in the previous section, but in a
flux background. Turning back to the complex equation
(\ref{0asol:cplx}),
we begin by rewriting it in terms of a {\it{complex}} dimensionless
variable
\ben
\Lambda=\frac{4n^2 t^2}{R^2} y^{-2(1-n/R)}=
\frac{4n^2 t^2}{R^2} y^{-2(1-n/R)} \, . \label{comp_lambda}
\een
This gives
\ben
\Lambda^{-\frac{1}{2(1-n/R)}}=e^{-2\tilde\chi}+e^{-\frac{2n}R(
\tilde\chi+\bar{\tilde\chi})}\bigg[\bigg(1-\frac nR\bigg)
e^{2\tilde\chi}-\frac nRe^{2\bar{\tilde\chi}}\bigg]\label{0a}
\een
where $\tilde\chi=\partial\phi+\ln(\frac{4n^2 t^2}{R^2})$ is the
complex reduced susceptibility.
Of course, we will also need the complex conjugate of equation
(\ref{0a}).

Let us pause to make a brief point alluded to in section
\ref{relevant}. Just as in the $q=0$ case, the relationship between the
dimensionless parameter \eqref{comp_lambda} and the unperturbed theory
becomes ill-defined for $R<n$. In the absence of flux, it is understood
that we must have $R>n$ so that the perturbation of the worldsheet
Lagrangian by ${\cal V}_{2n/R}$ is relevant. We have no such worldsheet
understanding with $q \ne 0$, but we still have a breakdown in our
variables. This suggests that the relevance bound $R>n$ may be extended
for the case with flux. Following this suggestion, we will only
consider $R>n$ in the following.

To make the notation more concise, we find it useful to introduce the
complex function
\bea
A=e^{2(1-\frac nR)\tilde\chi}e^{-\frac{2n}R\bar{\tilde\chi}}.
\eea
The complex equation (\ref{0a}) can be written in terms of $A,\bar A$ as
the system
\bea
\frac{A}{\bar A}&=&
\frac{\bar \Lambda^{-\frac{1}{2(1-n/R)}}-(1-\frac nR)\bar A
+\frac nRA}{\Lambda^{-\frac{1}{2(1-n/R)}}-(1-\frac nR)A+\frac nR\bar A} \, ,
\nonumber\\
\bigg(A\bar A\bigg)^{-\frac 1{1-2n/R}}
&=&\bigg|\Lambda^{-\frac{1}{2(1-n/R)}}-(1-\frac nR) A+\frac nR
\bar A\bigg|^2 \, . \label{A}
\eea

As in the $q=0$ case, we identify the critical points from the
divergences of the third order derivatives of the partition function,
or more precisely of the derivatives $\partial_\Lambda \tilde\chi$
and $\partial_\Lambda \bar{\tilde\chi}$.
Therefore one begins by taking  a $\Lambda$-derivative of (\ref{0a})
and its
complex conjugate
\bea
\frac{\Lambda^{-\frac{1}{2(1-n/R)}-1}}{4(1-\frac nR)^2}&=&\bigg[
\frac{\Lambda^{-\frac{1}{2(1-n/R)}}}{1-\frac nR}
-(2-\frac{n}R)A+\frac nR\bar A\bigg]
\partial_\Lambda \tilde\chi
+\frac nR \bigg(A+\bar A\bigg)
\partial_\Lambda \bar{\tilde\chi} \, ,\nonumber\\
0&=&\frac nR\bigg(A+\bar A\bigg)
\partial_\Lambda \tilde\chi+
\bigg[\frac{\bar\Lambda^{-\frac{1}{2(1-n/R)}}}{1-\frac nR}
-(2-\frac{n}R)\bar A+\frac nRA\bigg]
\partial_\Lambda \bar{\tilde\chi} \, .
\eea
This is a linear algebraic system for the two derivatives
$\partial_\Lambda\partial\tilde\phi$
and $\partial_\Lambda\bar\partial\tilde\phi$. Therefore any
divergent (critical) behavior will be associated with the zeros of the
discriminant
\bea
\bigg|\frac{\Lambda_c^{-\frac 1{2(1-n/R)}}}{1-\frac nR} -
(2-\frac nR)A_c +\frac nR \bar A_c\bigg|^2
-(\frac nR)^2(\bar A_c+A_c)^2=0 \, ,
\label{0a:crit}
\eea
where the subscripts are meant to imply that \eqref{0a:crit} only holds
on the critical surface.

To reiterate, the 0A critical surface is constrained by
(\ref{0a:crit}), where
$A,\bar A$ are defined through (\ref{A}). To obtain the critical
surface we must first
eliminate $A_c,\bar A_c$ in terms of $\Lambda_c$ and $R$.
We proceed by combining (\ref{0a:crit}) and the first equation in
(\ref{A}), and
find that we can solve for the critical $A_c,\bar A_c$ from a quartic
equation
\com{\bea
&&(2x-1) z_c^4+4 L_{1,c}(3x-2) z_c^3 + 24 L_{1,c}^2 (x-1) z_c^2 \nn \\
   &&~~~~~~~~~~~~~~~~~~~~~~~~~~~~~~~~~~   +16 L_{1,c}^3(x-2)z_c-16L_{1,c}^2( 
L_{1,c}1^2+L_{2,c}^2)=0 \, ,\label{quartic}
\eea}
\ben
(2x-1) z_c^4+4 L_{1,c}(3x-2) z_c^3 + 24 L_{1,c}^2 (x-1) z_c^2+16 
L_{1,c}^3(x-2)z_c-16L_{1,c}^2( L_{1,c}^2+L_{2,c}^2)=0 \, ,\label{quartic}
\een
where
\bea
z=4(x-1) Re(A)\, , \qquad L_1=Re(\Lambda^{-\frac1{2(1-n/R)}})\,
,\qquad
L_2=Im(\Lambda^{-\frac1{2(1-n/R)}})\, ,\qquad x=\frac nR\,.
\label{vardef}
\eea
The powers appearing in the definitions of $L_{1,2}$ are such that
$L_1\sim \mu$
and $L_2\sim q$ (with proportionality factors depending on $R,t$).

\begin{figure}[t]
\noindent 
\begin{centering}\includegraphics[width=0.4\textwidth,keepaspectratio]{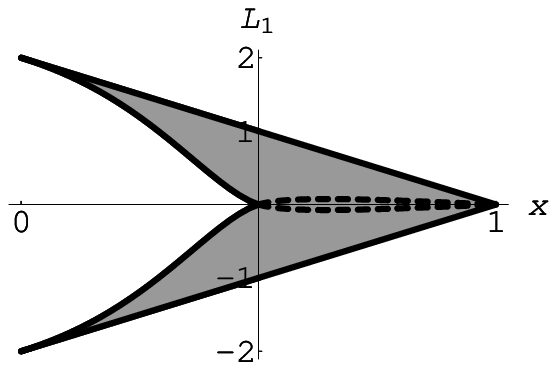}
\includegraphics[width=0.4\textwidth,keepaspectratio]{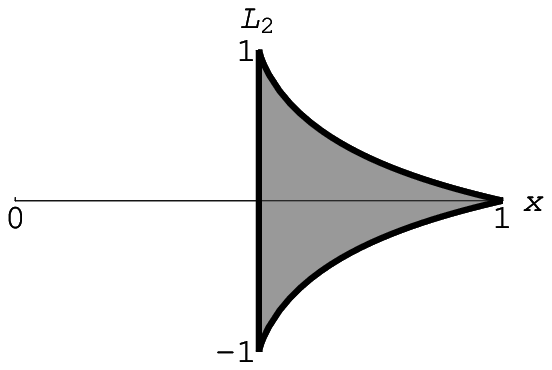}\par\end{centering}

\caption{\label{Plots4}\footnotesize{The left plot depicts a slice through the
0A critical surface with $L_2 \sim q=0$.
The inner curve $L_1=L_1(x)$ (closer to the $x$-axis) is the $c=1$ critical 
curve
discussed in section \ref{critq0}, where $\partial_\mu^2 \phi$ diverges, while 
the outer line
is a new feature of the 0A critical surface, where $\partial_q^2 \phi$ 
diverges. The dotted lines indicate solutions of \eqref{0asol} where 
$\partial_\mu^2 \phi$ diverges, but which are probably not accessible from the 
unperturbed 0A theory since they are completely enclosed by other critical 
surfaces.
The right plot takes $L_1=0$ and shows the
dependence on the flux parameter $L_2$.
}}
\end{figure}

\begin{figure}[t!]
    \begin{tabular}{cc}
\leavevmode
\!\!\!\!\!\!\!\!\!\!\!\!\!\!\!\!\!\!\!\!\epsfxsize 9cm
\epsfbox{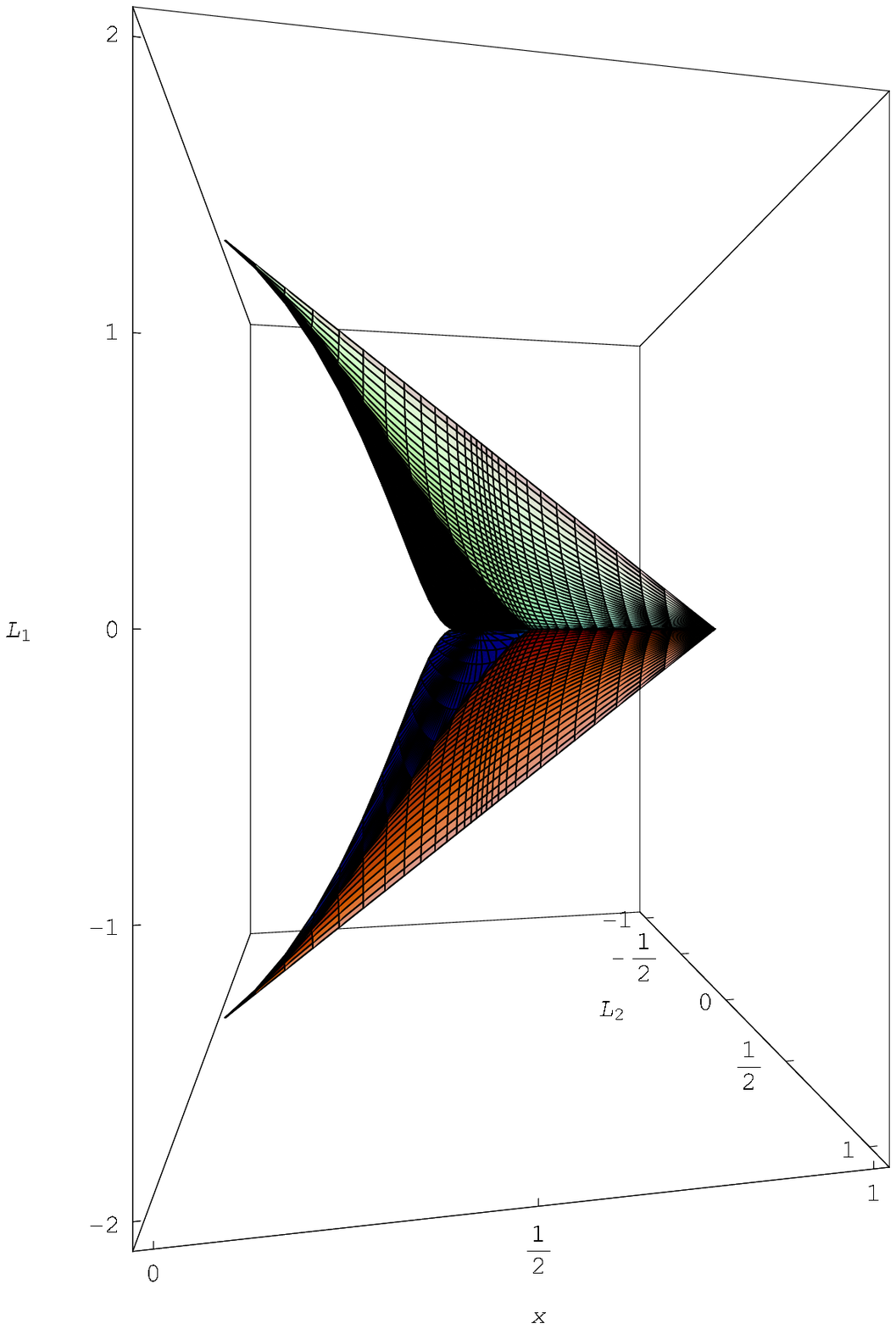}&
      \leavevmode
\epsfxsize 9cm \epsfbox{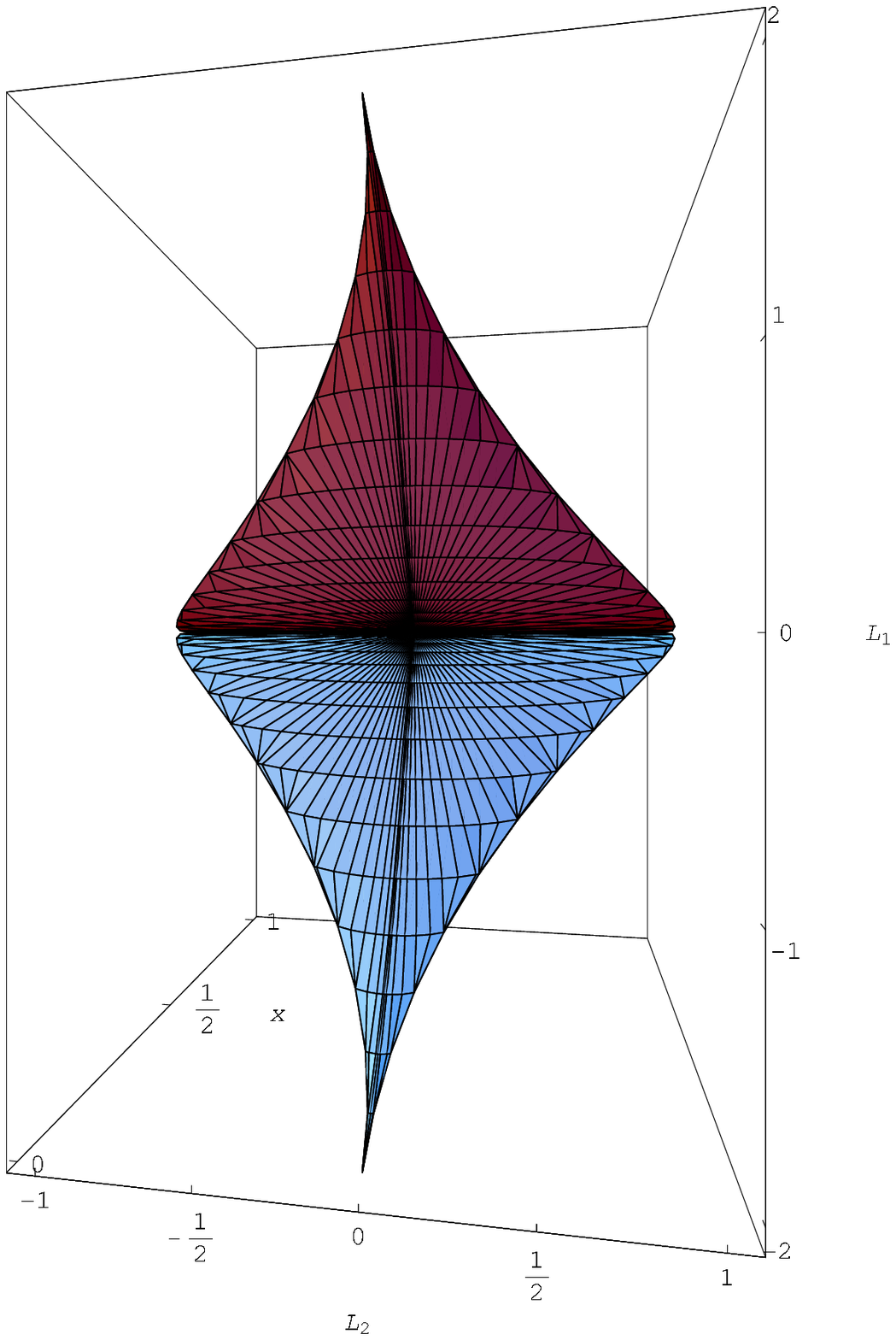} \\
    \end{tabular}
{\caption{\label{Plots5}\footnotesize{The left plot depicts the 0A
critical surface,
with emphasis on the dependence of $L_1$ on $x$. The right plot depicts
the
same surface, but from "below", which emphasizes the extent in
the $L_{1,2}$ plane. To get oriented, compare the left 3d surface with the left 
cross-section in Figure \eqref{Plots4}. The "tentacles" of the left surface 
continue all the way to $x=0$, where they collapse to the points $L_1 =\pm 2$, 
$L_2 =0$. At $x=1$, the surface collapses to a point at $L_1 =L_2 =0$.}}}
\end{figure}

Before studying the entire critical surface, let us consider the zero flux 
limit, $L_2=0$,
where we expect to recover the results of section \ref{critq0}.
In this case we find that the solutions to the quartic equation are
given by
$Re(A_c)=\bigg(\frac{2}{2x-1},-2,-2,-2\bigg) \frac{L_{1,c}}{4(x-1)}$.
The degeneracy observed among the roots in this special case is only
accidental.
The first (non-degenerate) root corresponds to the $c=1$ critical surface 
\eqref{ceq1critsurf}, as can be verified by comparison with \eqref{divdchi2}. 
By expanding (\ref{A}) around this non-degenerate root, while keeping
the RR flux zero, we can extract the divergent behaviour of 
$\partial_\mu^2 \phi\sim (L_1-L_{1,c})^{-1/2}\sim
\left(Re(\Lambda)-Re(\Lambda_c)\right)^{-1/2}$. 
This was expected, and it is in accord with (\ref{c=0}). 

The second (degenerate) root is novel, and corresponds to a straight line, 
$\left|L_{1}\right|=2\left(1-x\right)$, where $\partial_q^2 \phi$ diverges.
The critical behaviour at zero RR flux is given by
$\partial_q^2 \phi\sim (L_1-L_{1,c})^{-1}\sim 
(Re(\Lambda)-Re(\Lambda_c))^{-1}$.
This clearly distinguishes $c=1$ perturbed by momentum modes from 0A with 
$q=0$, and the same perturbation. In Figure \eqref{Plots4} we include plots of 
the intersection of the 0A critical surface with the $L_2 =0$ and $L_1 =0$ 
planes.

The full 0A critical surface is found by substituting the solutions of the 
quartic into the remaining
(second) equation of (\ref{A}). This constraint defines a surface in
the 3d
space parameterized by $L_1$, $L_2$ and $x$, which were introduced in
\eqref{vardef}.
We used Mathematica to generate the plots of the 0A critical surface shown in
Figure (\ref{Plots5}). We have plotted the
0A surface in terms of the finite range variable $x\in (0,1)$
(\textit{i.e.} $R>n$) and
as a function of the real and imaginary parts of the dual
``Sine-Liouville''
dimensionless parameter, $\Lambda^{-\frac{1}{2(1-n/R)}}$.
Note that $L_1^2 + L_2^2\to \infty$ is the regime identified with the 
unperturbed 0A string theory.
The entire 0A critical surface is symmetric with respect to the $L_2=0$
plane as well
as the $L_1=0$ plane. It is also closed.



The line $L_1=L_2=0$ of Figures (\ref{Plots4}) and \eqref{Plots5} is the 
momentum
mode condensate (Sine-Liouville) phase.
Fixing $L_2 =0$ (as in the left plot of Figure (\ref{Plots4})) it
would seem that the
critical surface obstructs a smooth connection to the unperturbed 0A
regime $|L_1|\gg 1$.
For $x\in (0,{1\over 2})$ the obstruction includes the familiar $c=1$
critical curve \eqref{ceq1critsurf}. The novel (linear) component of the 
critical curve
obstructs
the entire range of  $x\in (0,1)$. In Figure (\ref{Plots5}) we see that, for 
the
range $x\in (0,{1\over 2})$
the obstruction can be avoided by taking advantage of non-trivial flux
$L_2$, while
for $x\in ({1\over 2},1)$ the
Sine-Liouville regime
remains covered by the critical surface. 
Note that this is precisely the opposite of the case in $c=1$, where
$x\in (0,{1\over 2})$ is obstructed and  $x\in ({1\over 2},1)$ is not 
\cite{Kazakov:2000pm}.

\section{Outlook \& Holomorphic Perturbations}

There are a number of obvious extensions of the line of development presented 
in this work. We have studied the perturbation of 0A string theory by momentum 
mode operators in the presence of Ramond-Ramond flux. It would be of interest 
to address also the effect of perturbing by winding operators when flux is 
present; some prior work in this direction includes 
\cite{Yin:2003iv,Park:2004yc}. Although of more technical difficulty (since the 
matrix model is no longer in its singlet sector), such perturbations may shed 
light on the conjectured 0A black hole \cite{Maldacena:2005hi}. Also, the study 
of the integrable structure of 0B string theory with non-trivial flux is of 
obvious interest. The analysis of winding and momentum mode perturbations in 
that case would provide another probe of T-duality for Type 0 strings. All of 
these directions would provide a more detailed picture of the vacuum structure 
of two-dimensional string theory.

There is a more novel direction which is suggested by the complexified Toda 
lattice that we have introduced. We have not fully explored this structure for 
we have only examined deformations by the momentum modes, which are given by 
$\hat{B}_\pm \sim e^{2i\partial_\epsilon}$. Formally, it is natural to further 
exploit the complex nature of the system by studying perturbations which 
preserve the holomorphicity existing at vanishing perturbation 
\cite{Maldacena:2005he}. Such perturbations involve the flux in a non-trivial 
way.

To make this more concrete, we consider the transformations
\ben
\hat{U}_{\pm}=e^{i\sum_{n>0}(\tilde{b}_{\pm 
n}\left(\hat{y};\{t\}\right)\hat{z}_{\pm}^{-n/R} +{\rm h.c.}) }e^{\mp 
i\lt(\tilde{\phi}\left(\hat{y};\{t\}\right)/2+{\rm h.c.}\rt)}e^{i\sum_{n>0} 
(t_{\pm n}\hat{z}_{\pm}^{n/R}+{\rm h.c.})}, \label{Uholo}
\een
for some complex parameters $\{t\}$ and undetermined functions $\tilde{b}$ and 
$\tilde{\phi}$ and where ``h.c.'' denotes Hermitian conjugate. We will call 
these ``holomorphic perturbations'' since \eqref{Uholo} factorizes into an 
operator dependent only on $\hat{y}$ and $\hat{\eta}^{\pm 1}$ and one dependent 
only on $\hat{\bar{y}}$ and $\hat{\bar{\eta}}^{\pm 1}$. Thus, this is much like 
two decoupled copies of the transformation \eqref{eq:u-dress} for the $c=1$ 
matrix model.

The string equation \eqref{0Astring} is the same under this transformation as 
when perturbed by momentum modes. The expansions of the complex Orlov-Shulman 
operator in terms of the Lax operators are now
\bea
\hat{Y}_{+} & = & \hat{y}-\frac{2n}{R} \bar{t}_{+ n}\hat{\bar{Z}}_{+}^{\, 
n/R}+\sum_{k>0}\bar{v}_{+ k}\left(\hat{y}\right) \hat{\bar{Z}}_{+}^{\, -k/R}, 
\\
\hat{Y}_{-} & = & \hat{y} + \frac{2n}{R} t_{- n}\hat{Z}_{-}^{\, 
n/R}+\sum_{k>0}v_{- k}\left(\hat{y}\right)\hat{Z}_{-}^{\, -k/R},
\eea
for some undetermined functions $v_{\pm k}$. Combining the above expansions 
with the string equation we obtain the holomorphic susceptibility equation
\ben
y-e^{-2\partial\phi} = 
\left(\frac{2n}{R}\right)^{2}t_{-n}\bar{t}_{+n}e^{2\left(1-n/R\right)\partial\phi}\left(\frac{n}{R}-1\right).
\een
Unlike \eqref{0asol:cplx}, the above result does not mix $\partial \phi$ and 
$\bar{\partial} \phi$ and so preserves the holomorphic factorization of the 
unperturbed solution.

It is important to keep in mind that the physical interpretation of the 
holomorphic perturbations is quite different from that of the momentum mode 
perturbations studied throughout this work. In the absence of flux, perturbing 
by the matrix model operators $\hat{B}_\pm$ has the clear interpretation of 
adding tachyon vertex operators to the string worldsheet action; this can be 
generalized to a flux background since $\hat{B}_\pm$ still generate small 
deformations in the Fermi sea at finite $q$. The holomorphic perturbations 
above are less clearly interpreted, and perhaps should be understood as 
generating coherent states of D-branes or discrete states. Alternatively, the 
interpretation may lie outside of 0A string theory itself and require ideas 
from non-critical M-theory \cite{Horava:2005tt,Horava:2005wm}. It would be of 
great interest to understand the full physical interpretation of these 
perturbations and the complex Toda structure introduced herein.

\section*{Acknowledgements}

It is a pleasure to thank P. de Medeiros, P. Ho\v rava, J. Maldacena and L. 
Pando Zayas for useful discussions. This work was supported by the DOE.

\appendix

\section{Chiral Quantization of $c=1$ Matrix Model}\label{c1chiral}

Here we will briefly review the chiral quantization of the fermion system dual 
to $d=2$ bosonic string theory. Since these fermions are non-interacting we 
will make do with single-particle quantization.

\subsection{Operator algebra}

Recall that the single-particle energy for the fermions in the $c=1$ matrix 
model is given by

\ben
\hat{\epsilon}=\frac{1}{2}\left(\hat{p}^{2}-\hat{x}^{2}\right).
\een
The algebraic structure becomes clearer when we introduce the light-cone 
coordinates
\ben
\hat{x}_{\pm}\equiv\frac{1}{\sqrt{2}}\left(\hat{p}\pm\hat{x}\right),
\een
which obey the following closed algebra (note $\hbar =1$)
\begin{eqnarray}
\left[\hat{x}_{+},\hat{x}_{-}\right] & = & i, \nn \\
\left[\hat{x}_{\pm},\hat{\epsilon}\right] & = & \pm i\hat{x}_{\pm}, \nn \\
\left\{ \hat{x}_{+},\hat{x}_{-}\right\}  & = & 2 \hat{\epsilon}. 
\label{eq:c=1Comm}
\end{eqnarray}
The first commutator indicates that $\hat{x}_\pm$ are canonically conjugate, 
and the second that $\hat{x}_\pm$ are ``shift operators'' of $\hat\epsilon$. 
The third relation indicates that the Schrodinger equation is first order in 
both the $x_\pm$ bases, which is a sizable advantage over the second order wave 
equation in the position basis.

Using the above relations the Schrodinger equation can be expressed in either 
the $x_\pm$ bases
\ben
\mp i \left( x_{\pm} \partial_{x_{\pm}} + \half \right) \Psi (x_\pm,t) = i 
\partial_t \Psi (x_\pm,t).
\een
Or for energy eigenfunctions, where $\Psi_\epsilon (x_\pm,t) = e^{-i \epsilon 
t} \psi_\epsilon (x_\pm)$,
\ben
\left( x_{\pm} \partial_{x_{\pm}} + \half \right) \psi_\epsilon (x_\pm) = \pm i 
\epsilon \, \psi_\epsilon (x_\pm).
\een

\subsection{States and eigenfunctions}

For a given energy, $\epsilon$, there are four states of interest which in ket 
notation are $| \, \epsilon, in/out, L/R \R$. The notation is such that $in$ 
refers to fermions moving toward the origin and $out$ refers to those moving 
away. The label $L/R$ refers to which side of the potential the fermion is 
localized on. The $in$ ($out$) eigenfunctions are most conveniently expressed 
in the $x_-$ ($x_+$) bases. They are as follows
\bea
\label{eigen}
\L x_- | \, \epsilon, in, R \R & = &  \theta(x_-) \frac{x_-^{-i \epsilon - 
\half}}{\sqrt{2 \pi}}, \nn \\
\L x_- | \, \epsilon, in, L \R & = &  \theta(-x_-) \frac{(-x_-)^{-i \epsilon - 
\half}}{\sqrt{2 \pi}}, \nn \\
\L x_+ | \, \epsilon, out, R \R  & = & \theta(x_+) \frac{x_+^{i \epsilon - 
\half}}{\sqrt{2 \pi}}, \nn \\
\L x_+ | \, \epsilon, out, L \R  & = & \theta(-x_+) \frac{(-x_+)^{i \epsilon - 
\half} }{\sqrt{2 \pi}}.
\eea
The \textit{in} and \textit{out} states are not independent, but rather are 
connected by the change of basis, $x_- \leftrightarrow x_+$.

We can see that the $in$ eigenfunctions are delta-function normalized by 
inserting a complete set of $x_-$ eigenstates.
\bea
\L \, \epsilon^\prime, in , a | \, \epsilon, in, b \R & = & 
\int_{-\infty}^{\infty} \, dx_- \L \, \epsilon^\prime , in, a | x_- \R \L x_- | 
\, \epsilon, in, b \R \nn \\
& = & \delta(\epsilon^\prime - \epsilon) \delta_{a,b},
\label{innorm}
\eea
\noindent where the Roman letters label $L/R$. Similarly, by inserting a 
complete set of $x_+$ states, we can show
\ben
\label{outnorm}
\L \, \epsilon^\prime, out , a | \, \epsilon, out, b \R = 
\delta(\epsilon^\prime - \epsilon) \delta_{a,b}.
\een
It is also useful to consider parity eigenstates. Define
\bea
\label{parity}
| \, \epsilon, in, \pm \R & = & \frac{1}{\sqrt{2}} \left(| \, \epsilon, in, R 
\R \pm | \, \epsilon, in, L \R \right),   \nn \\
| \, \epsilon, out, \pm \R & = & \frac{1}{\sqrt{2}} \left(| \, \epsilon, out, R 
\R \pm | \, \epsilon, out, L \R \right),
\eea
where the third quantum number labels the parity eigenvalue.

\subsection{Unperturbed Scattering}

Scattering in the unperturbed problem involves only a simple change of basis. 
We want to compute $\L \, \epsilon^\prime, out , a | \, \epsilon, in, b \R$. 
This is easily done by inserting complete sets of states of both $x_{\pm}$. 
Note that in performing the calculations, one must use $\L x_+ | x_- \R = 
\frac{1}{\sqrt{2\pi}} e^{i x_+ x_-}$ which follows from the canonical 
commutator $[x_+,x_-] = i$. We will also need use of the integrals 
\cite{Gradshteyn}
\bea
\Gamma (z) &=& \frac{b^z}{\sin \frac{\pi z}{2}} \int_0^\infty dt\, 
\sin\lt(bt\rt) t^{z-1} \nn \\
\Gamma (z) &=& \frac{b^z}{\cos \frac{\pi z}{2}} \int_0^\infty dt\, 
\cos\lt(bt\rt) t^{z-1}.
\eea
The scattering is completely diagonal in the parity basis
\bea
\L \, \epsilon^\prime, out , \pm | \, \epsilon, in, \mp \R & = & 0, \nn \\
\L \, \epsilon^\prime, out , \pm | \, \epsilon, in, \pm \R & = & 
\frac{1}{\sqrt{2  \pi} }\lt( e^{{i \pi \over 4} +{\epsilon \pi \over 2}} \pm 
e^{-{i \pi \over 4} -{\epsilon \pi \over 2}}\rt) \Gamma\lt({1 \over 2} - i 
\epsilon \rt) \delta(\epsilon^\prime -\epsilon) \nn \\
    & = & e^{i \pi /4} \sqrt{1 \mp i e^{-\pi\epsilon} \over 1 \pm i 
e^{-\pi\epsilon}} \sqrt{ \Gamma \lt(\half - i \epsilon \rt) \over \Gamma 
\lt(\half + i \epsilon \rt) } \delta(\epsilon^\prime -\epsilon).
\eea
To discuss the $c=1$ matrix model we consider only the parity odd states. Then 
we can write the scattering as
\ben
| \, \epsilon, in, - \R = S(\epsilon) | \, \epsilon, out , - \R, \label{keteq}
\een
where the bounce factor is defined as
\bea
S(\epsilon) & = & e^{i \pi /4} \sqrt{1 + i e^{-\pi\epsilon} \over 1 - i 
e^{-\pi\epsilon}} \sqrt{ \Gamma \lt(\half - i \epsilon \rt) \over \Gamma 
\lt(\half + i \epsilon \rt) } \nn \\
            & \approx & e^{-i \pi /4} \sqrt{ \Gamma \lt(\half - i \epsilon \rt) 
\over \Gamma \lt(\half + i\epsilon\rt)}. \label{bounce}
\eea
In the second line we have neglected terms of order $e^{\pi \epsilon}$, {\it 
i.e.} we consider large negative energy with negligible tunneling. This is 
appropriate and indeed required in the $c=1$ model which is defined only by 
perturbation theory in $\frac{1}{|\epsilon|}$.

The expression (\ref{keteq}) indicates the \textit{in} and \textit{out} states 
are not independent. Indeed, \eqref{keteq} appears to be the conventional 
relationship between scattered states, but this is incorrect. Both $ | \, 
\epsilon, in/out , - \R$ are solutions to the time-independent Schrodinger 
equation and so \eqref{keteq} is an equality valid at {\it all finite times}. 
That is, there is no implicit insertion of the time translation operator 
evolving time from $t=-\infty$ to $t=+\infty$. To eliminate this redundancy, 
define
\ben
| \, \epsilon \R = S^{-1/2} (\epsilon)   | \, \epsilon, in, - \R = S^{1/2} 
(\epsilon) | \, \epsilon, out , - \R .
\een
Then we have two equivalent expressions for the same state, given in different 
bases
\ben
\psi_\epsilon^\pm (x_\pm) = \L \, x_\pm | \, \epsilon \R = \frac{S^{\pm 1/2} 
(\epsilon)}{\sqrt{4\pi}} sign(x_\pm) |x_\pm |^{\pm i \epsilon -\half}. 
\label{normalized}
\een
Even though we consider the $c=1$ theory where the fermions are localized on 
one side of the barrier, to calculate probabilities one should still integrate 
over all $x_\pm$, \textit{i.e.} the wavefunctions (\ref{normalized}) are 
delta-function normalized on the whole interval $x_\pm \in \mathbb{R}$.

\section{Chiral Quantization of 0A Matrix Model}\label{sub:0APhase}

We reproduce here results from the Appendix of \cite{Maldacena:2005he} for our 
normalization differs in its $q$-dependence. Consider now particles moving in a 
plane with a radial harmonic oscillator potential which in Cartesian 
coordinates has the Hamiltonian
\ben
\hat{\epsilon} = \half \lt( \hat{p}_1^2 - \hat{x}_1^2 + \hat{p}_2^2 - 
\hat{x}_2^2 \rt)\,.
\een
It is useful to apply chiral quantization as in $c=1$ to both degrees of 
freedom,
\bea
\hat{x}_{\pm,i} & \equiv & \frac{\hat{p}_{i}\pm \hat{x}_{i}}{\sqrt{2}}\,,\nn\\
\left[\hat{x}_{\pm,i}, \hat{x}_{\mp,j}\right] & = & \pm i\delta_{ij}\,.
\eea
For future use, note that the kernel of this transformation is
\begin{equation}
\left\langle \left.x_i \right|x_{\pm , i}\right\rangle =\frac{2^{1/4}e^{\pm 
i\pi/8}}{\sqrt{2\pi}}\exp\left[\mp i\left(\frac{x_i^{2}}{2}\mp\sqrt{2}x_i 
x_{\pm , i} + \frac{x_{\pm , i}^{2}}{2}\right)\right] \, 
,\label{eq:LightConeConv}
\end{equation}
where we have chosen the phase in \eqref{eq:LightConeConv} so as to have
a simple form for the inner-product
\ben
\left\langle \left.x_{+,i}\right|x_{-,i}\right\rangle 
=\frac{1}{\sqrt{2\pi}}\exp \left( ix_{+,i}x_{-,i}\right) \,.
\een

It is not as straightforward to compute the relevant inner products and 
wave-functions as in the $c=1$ theory so we must go through a series of changes 
of variables. First, introduce the conventional polar coordinates
\ben
\hat{x}_1  =   \hat{r} \cos \hat\theta\, , \qquad \hat{x}_2  =  \hat{r} \sin 
\hat\theta \, .
\een
So that $| x_1 , x_2 \R$ are delta-function normalized, the kernel of this 
change of basis is
\ben
\left\langle \left.x_{1},x_{2}\right|r,\theta \right\rangle 
=\sqrt{x}\delta\left(x_{1}-r\cos\theta\right)\delta\left(x_{2}-r\sin\theta\right) 
\, .\label{polarkernel}.
\een
The momentum conjugate to $\theta$ is the conserved charge $\hat{q} = \hat{x}_1 
\hat{p}_2 - \hat{x}_2 \hat{p}_1$.
It is also of use to introduce polar coordinates in phase space via
\ben
\hat{x}_{\pm, 1}  =   \hat{r}_\pm \cos \hat\theta_\pm\, , \qquad 
\hat{x}_{\pm,2}  =  \hat{r}_\pm \sin \hat\theta_\pm \, . \label{phasepolar}
\een
which have an integral kernel analogous to (\ref{polarkernel}). The relation 
between conventional polar coordinates and the phase space polar coordinates is 
easily derived to be
\begin{eqnarray}
\left\langle \left.r_{\pm},\theta_{\pm}\right|r,\theta\right\rangle  & = & \int 
dx_{1}dx_{2}dx_{\pm,1}dx_{\pm,2}\left\langle 
\left.r_{\pm},\theta_{\pm}\right|x_{\pm,1},x_{\pm,2}\right\rangle \nn\\
&  & \quad\quad\quad\quad\quad\times\left\langle 
\left.x_{\pm,1},x_{\pm,2}\right|x_{1},x_{2}\right\rangle \left\langle 
\left.x_{1},x_{2}\right|r,\theta\right\rangle \nn\\
& = & \frac{e^{\mp i\pi/4}}{2\pi}\sqrt{2r r_{\pm}}\exp\left[\pm 
i\left(\frac{r^{2}}{2}\mp\sqrt{2}r 
r_{\pm}\cos\left(\theta-\theta_{\pm}\right)+\frac{r_{\pm}^{2}}{2}\right)\right] 
\, .
\end{eqnarray}
It is convenient to exchange
$\theta_{\pm}$ for the conjugate momenta $q_{\pm}$
\begin{eqnarray}
\left\langle r,\theta\left|r_{\pm},q_{\pm}\right.\right\rangle  & = & 
\frac{e^{\pm i\pi/4}}{\left(2\pi\right)^{3/2}} \sqrt{2rr_{\pm}}e^{\mp i ( 
\frac{r^{2}}{2}+\frac{r_{\pm}^{2}}{2})}\int_{0}^{2\pi}d\theta 
e^{i\left(q_{\pm}\theta_{\pm}+\sqrt{2}rr_{\pm}\cos\left(\theta-\theta_{\pm}\right)\right)}\nn\\
& = & \frac{e^{\pm 
i\pi/4}i^{\left|q_{\pm}\right|}}{\sqrt{2\pi}}\sqrt{2rr_{\pm}}e^{\mp i 
(\frac{r^{2}}{2}+\frac{r_{\pm}^{2}}{2}\mp q_{\pm}\theta ) 
}J_{\left|q_{\pm}\right|}\left(\sqrt{2}rr_{\pm}\right),
\end{eqnarray}
and then calculate
\begin{eqnarray}
\left\langle r_{+},q_{+}\left|r_{-},q_{-}\right.\right\rangle  & = & 
\frac{e^{-i\pi/2}i^{\left|q_{-}\right|-\left|q_{+}\right|}}{2\pi}2\sqrt{r_{+}r_{-}}e^{i\left(r_{+}^{2}+r_{-}^{2}\right)/2}
\int_{0}^{2\pi}d\theta e^{i\theta\left(q_{-}-q_{+}\right)}\nonumber \\
&  & 
\times\int_{0}^{\infty}rdre^{ir^{2}}J_{\left|q_{+}\right|}\left(\sqrt{2}rr_{+}\right)J_{\left|q_{-}\right|} 
\left(\sqrt{2}rr_{-}\right)\nn\\
& = & 
\delta_{q_{+}q_{-}}i^{\left|q_{+}\right|}\sqrt{r_{+}r_{-}}J_{\left|q_{+}\right|}\left(r_{+}r_{-}\right) 
\, .
\end{eqnarray}

We almost have all of the ingredients for the scattering phase. First we note 
that the momenta conjugate to $\hat{r}_\pm$ must classically satisfy the 
Poisson brackets $\{ r_\pm,p_\pm \}=1$. Using $x_{+ , i}$ as the generalized 
coordinates and $x_{-,i}$ as their momenta, we easily obtain $p_\pm = \pm 
\epsilon / r_\pm$. When promoted to operators this is written
\ben
\hat\epsilon  =  \pm 
\frac{1}{2}\left(\hat{r}_{\pm}\hat{p}_{\pm}+\hat{p}_{\pm}\hat{r}_{\pm}\right) 
\, ,
\een
which yields the time-independent wavefunctions
\ben
\left\langle \left.r_{\mp}\right|\epsilon,in/out\right\rangle 
=\frac{1}{\sqrt{2\pi}}r_{\mp}^{\mp i\epsilon-1/2} .
\een
Finally, we compute the scattering phase by combining the results of this 
appendix,
\begin{eqnarray}
\left\langle \left.\epsilon^{\prime},q^{\prime},out\right| 
\epsilon,q,in\right\rangle & = &
\int_{0}^{\infty}dr_{-}dr_{+}\sum_{q_{+},q_{-}}\left\langle 
\left.\epsilon^{\prime},q^{\prime},out\right|r_{+},q_{+}\right\rangle
\left\langle \left.r_{+},q_{+}\right|r_{-},q_{-}\right\rangle \left\langle 
\left.r_{-}q_{-}\right|\epsilon,q,in\right\rangle  \nn \\
& = & 
\delta_{qq^{\prime}}\delta\left(\epsilon-\epsilon^{\prime}\right)2^{-i\epsilon}i^{\left|q\right|} 
\frac{\Gamma\lt(\frac{1}{2}\lt(1+\left|q\rt|-i\epsilon\rt)\rt)}{\Gamma\lt(\frac{1}{2}\lt(1+\lt|q\rt|+i\epsilon\rt)\rt)} 
\, .
\end{eqnarray}
Thus the scattering phase is given by
\ben
S(\epsilon,q) = 2^{-i\epsilon}i^{\left|q\right|} 
\frac{\Gamma\lt(\frac{1}{2}\lt(1+\lt|q\rt|-i\epsilon\rt)\rt)}{\Gamma\lt(\frac{1}{2}\lt(1+\lt|q\rt|+i\epsilon\rt)\rt)} 
\, .
\een

\section{An Alternate Derivation of the 0A String Equations}

Consider the compactified, unperturbed, all-genus 0A partition function 
\cite{Maldacena:2005he}
\ben
\ln {\cal Z} = -\half Re \int_0^\infty\frac{dt}t \frac{e^{itR(\mu +iq)/2}}{
\sinh(\frac{t}{4R}) \sinh (t/2)}, \qquad
\label{partition}
\een
where we have neglected terms analytic in $\mu$. Using the notation introduced 
previously, $y=\mu+iq$, $\partial= \frac{\partial}{\partial y}$, we arrive at 
the
functional equation
\bea
-4 \sin(\partial/(2R))\sin(\partial)\ln {\cal Z}=\ln(y)\label{func:holo}
\eea
where we have implicitly regularized the rhs of the above equation.
Similarly, we find
\bea
-4 \sin(\bar\partial/(2R))\sin(\bar\partial)\ln {\cal Z}=\ln(\bar y)
\eea

Following Kostov \cite{Kostov:2001wv}, we will map these functional equations
for the unperturbed partition function into the complex string equation derived
in Section 3. First, we shall identify the partition function with the
$\tau(y,\bar y)$-function of the complexified Toda hierarchy: $\ln {\cal Z}=
\ln(\tau)$.

Next, we recall that in the absence of perturbations, the Lax operators are
simply dressed complex shift operators
\bea
Z_- =W_- \eta^{- 1} W^{-1}_-,
\qquad \bar Z_+= W_+  \eta W_+^{-1}, \qquad {\rm plus \; the\; c.c.},
\eea
where the dressing functions, with the perturbations turned off are
$W_\pm = e^{\mp i\phi_0/2}$.
Moreover, given that the scattering phase and the $\tau$-function
are related through
\bea
\phi(\mu;q)=i\ln\bigg(\frac{\tau(\mu-\frac i{2R})}{\tau(\mu+\frac{i}{2R})}
\bigg)\,,
\eea
we derive
\bea
\frac{W_-(\mu)}{ W_+(\mu)}=\frac{\tau(\mu+\frac i{2R})}{\tau(\mu-\frac i{2R})}
\label{u+/u-}
\eea
which in terms of the complex variable $y$ reads
\bea
\frac{W_-(y,\bar y)}{W_+(y,\bar y)}=
\frac{\tau(y+\frac i{2R},\bar y+\frac{i}{2R})}{\tau(y-\frac i{2R},\bar y -
\frac{i}{2R})}
\eea
At this moment it is useful to also recall that the unperturbed
partition function,
and therefore $\tau$, enjoys a holomorphic factorization, already transparent
from (\ref{partition}).
This fact is a manifestation of the expression for the scattering phase, which
factorizes holomorphically in a manifest way. Clearly, this property
carries through
to the unperturbed dressing operators.
We conclude then that the previous equation can be decomposed
(up to normalization constants) in
\bea
\frac{W_-(y)}{W_+(y)}=
\frac{\tau(y+\frac i{2R})}{\tau(y-\frac i{2R})}\label{u+/u-holo}
\eea
and its complex conjugate.

We have now gathered all the information we need to proceed with
the derivation of the string equations. To this end we now turn back to the
functional-differential equation (\ref{func:holo}), which when integrated
yields
\bea
y=\frac{\tau(y+\frac {i}{2R}+i,\bar y)}{\tau(y-\frac{i}{2R}+i,\bar y)}
\frac{\tau(y-\frac{i}{2R}-i,\bar y)}{\tau(y+\frac i{2R}-i,\bar y)}
\eea
The latter expression, according to the previous discussion on the
holomorphic factorization
properties of the unperturbed $\tau$-function, reduces to
\bea
y=\frac{\tau(y+\frac {i}{2R}+i)}{\tau(y-\frac{i}{2R}+i)}
\frac{\tau(y-\frac{i}{2R}-i)}{\tau(y+\frac i{2R}-i)}
\eea
In the next step we substitute (\ref{u+/u-holo}), to arrive at
\bea
y=\frac{W_-(y+i)}{W_+(y+i)}
\frac{W_+(y-i)}{W_-(y-i)} \, .
\eea
Equivalently, this equation can be written as
\bea
&&\bigg[W_+(y)^{-1} W_-(y)\bigg]\,\eta^{-1}\,
\bigg[W_-(y)^{-1} W_+(y)\bigg]
\,\eta=y-i\nonumber\\
&&\eta \bigg[W_+(y)^{-1} W_-(y)\bigg]\,
\eta^{-1} \,\bigg[W_-(y)^{-1} W_+(y)\bigg]
=y+i\nonumber\\\label{pre-string}
\eea
Also, since $(W_-)^{-1} W_+ = e^{i\phi_0}$ it is a trivial statement
that $(W_-)^{-1} W_+ y [(W_-)^{-1} W_+]^{-1}=y$.
Furthermore, based on the Toda hierarchy flow equations, it can be
shown that the operator $(W_-)^{-1} W_+$ is independent of the couplings
$t_n, t_{-n}$. This means that (\ref{pre-string}) hold in general.
In fact, these equations constitute the constraint among the Lax and
Orlov-Shulman operators that we called the {\it string equations}
\bea
\bar Z_+ Z_-=Y+i\, ,\qquad
Z_-\bar Z_+=Y-i \, .
\eea
As highlighted by this derivation, the complex nature of the 0A string 
equations arises naturally from the holomorphicity of the unperturbed partition 
function.

\section{Comparison with Low-order Correlators}\label{app:correlators}

We here confirm the consistency of the complex susceptibility equation 
\eqref{0asol} with calculations of low-order tachyon correlators as stated in 
section \ref{pert_check}. To facilitate comparison with 
\cite{PandoZayas:2005tu} we first introduce
\begin{equation}
p\equiv\frac{2n}{R},
\end{equation}
the momentum of our perturbations, so that \eqref{0aradeq} becomes
\begin{equation}
e^{-2\partial_{\mu}\phi}=\frac{\mu^{2}}{\left[1-\left(p-1\right)p^{2}t^{2}e^{-\left(p-2\right)\partial_{\mu}\phi}\right]^{2}}+\frac{q^{2}}{\left[1-p^{2}t^{2}e^{-\left(p-2\right)\partial_{\mu}\phi}\right]^{2}}.
\end{equation}
We then expand $\partial_{\mu}\phi$ as
\begin{equation}
-\partial_{\mu}\phi=\frac{1}{2}\log\left(\mu^{2}+q^{2}\right)+\sum_{m>0}\frac{t^{2m}}{\left(m!\right)^{2}}a_{m}, 
\label{dphi_expand}
\end{equation}
and solve iteratively for the $a_{m}.$ Recalling \eqref{dphi_generate}
\begin{equation}
\partial_{\mu}\phi=R^{-1}\partial_{\mu}^{2}\log\mathcal{Z}=R^{-1}\left[\partial_{\mu}^{2}\log\mathcal{Z}_{0}+\sum_{m>0}\frac{t^{2m}}{\left(m!\right)^{2}}\partial_{\mu}^{2}\left\langle 
\mathcal{T}_{2n/R}^{m}\mathcal{T}_{-2n/R}^{m}\right\rangle \right],
\end{equation}
we can identify the coefficients in \eqref{dphi_expand} with the tachyon 
correlators
\begin{equation}
a_{m}=-R\partial_{\mu}^{2}\left\langle 
\mathcal{T}_{p}^{m}\mathcal{T}_{-p}^{m}\right\rangle .
\end{equation}
We reproduce the first few correlators here, as predicted by 
\eqref{dphi_expand}:
\begin{eqnarray}
\partial_{\mu}^{2}\left\langle \mathcal{T}_{p}\mathcal{T}_{-p}\right\rangle  & 
= & 
-R^{-1}p^{2}\left(\mu^{2}+q^{2}\right)^{\left(p-4\right)/2}\left[\left(p-1\right)\mu^{2}+q^{2}\right],\\
\partial_{\mu}^{2}\left\langle 
\mathcal{T}_{p}^{2}\mathcal{T}_{-p}^{2}\right\rangle  & = & 
-2R^{-1}\left[p^{2}\left(\mu^{2}+q^{2}\right)^{\left(p-4\right)/2}\right]^{2}\left[\left(p-1\right)^{2}\left(2p-3\right)\mu^{4}\right.\nonumber 
\\
&  & 
\left.+\left(18-22p+7p^{2}\right)\mu^{2}q^{2}+\left(2p-3\right)q^{4}\right],\\
\partial_{\mu}^{2}\left\langle 
\mathcal{T}_{p}^{3}\mathcal{T}_{-p}^{3}\right\rangle  & = & 
-6R^{-1}\left[p^{2}\left(\mu^{2}+q^{2}\right)^{\left(p-4\right)/2}\right]^{3}\left[\left(3p-5\right)\left(3p-4\right)\left(p-1\right)^{3}\mu^{6}\right.\nonumber 
\\
&  & +3\left(100-247p+239p^{2}-106p^{3}+18p^{4}\right)\mu^{4}q^{2}\nonumber \\
&  & 
+\left.3\left(-100+179p-108p^{2}+22p^{3}\right)\mu^{2}q^{4}+\left(3p-5\right)\left(3p-4\right)q^{6}\right].
\end{eqnarray}

We now wish to compare with the results of \cite{PandoZayas:2005tu}, which 
quotes several $A_{m}$ (see their (3.11)\footnote{For the reader's convenience, 
we point out that in \cite{PandoZayas:2005tu}
$a\equiv1$ and $f\equiv q/\mu.$ }) defined through
\begin{equation}
\left\langle 
\mathcal{T}_{0}\mathcal{T}_{p}^{m}\mathcal{T}_{-p}^{m}\right\rangle 
=\left(\mu^{2}+q^{2}\right)^{mp/2}A_{m}.
\end{equation}
These correlators $\left\langle 
\mathcal{T}_{0}\mathcal{T}_{p}^{m}\mathcal{T}_{-p}^{m}\right\rangle ,$
are related to $\left\langle 
\mathcal{T}_{p}^{m}\mathcal{T}_{-p}^{m}\right\rangle$, in both $c=1$ and 0A, 
via
\begin{equation}
\partial_{\mu}\left\langle \mathcal{T}_{p}^{m}\mathcal{T}_{-p}^{m}\right\rangle 
=\left\langle 
\mathcal{T}_{0}\mathcal{T}_{p}^{m}\mathcal{T}_{-p}^{m}\right\rangle .
\end{equation}
This was first recorded in \cite{Moore:1991zv} and exploited in 
\cite{Moore:1992ga}. We thus obtain
\begin{equation}
\partial_{\mu}^{2}\left\langle 
\mathcal{T}_{p}^{m}\mathcal{T}_{-p}^{m}\right\rangle 
=\partial_{\mu}\left[\left(\mu^{2}+q^{2}\right)^{mp/2}A_{m}\right].
\end{equation}
Using this expression and the $A_{m}$ in \cite{PandoZayas:2005tu} we see that 
our correlators match, up to an overall normalization factor $R$.

\bibliographystyle{utcaps}
\bibliography{2DStrings}

\end{document}